\documentclass[sigplan,screen]{acmart}
\usepackage{graphicx}
\usepackage{booktabs}
\usepackage{tabularx}
\usepackage{makecell}
\usepackage{subcaption}
\usepackage{pifont}
\usepackage{microtype}
\usepackage{multirow}
\usepackage{xcolor}

\newcounter{takeaway}
\newenvironment{takeawaybox}[1][]{%
  \refstepcounter{takeaway}%
  \ifx\\#1\\\else\label{#1}\fi
  \par\noindent
  \begin{tabular*}{\columnwidth}{@{}p{\columnwidth}@{}}
    \toprule
    \textbf{Takeaway \thetakeaway.} \itshape\ignorespaces
}{%
    \\ \bottomrule
  \end{tabular*}
  \par
}

\AtBeginDocument{%
  }

\copyrightyear{2027}
\acmYear{2027}
\setcopyright{cc}
\setcctype{by}
\acmConference[ASPLOS '27]{Proceedings of the 32nd ACM International Conference on Architectural Support for Programming Languages and Operating Systems, Volume 1}{April 11--15, 2027}{Heraklion, Greece}
\acmBooktitle{Proceedings of the 32nd ACM International Conference on Architectural Support for Programming Languages and Operating Systems, Volume 1 (ASPLOS '27), April 11--15, 2027, Heraklion, Greece}
\acmDOI{10.1145/3845814.3850089}
\acmISBN{979-8-4007-3012-2/2027/04}

\makeatletter
\let\arxiv@copyrightfootnote\footnotetextcopyrightpermission
\renewcommand\footnotetextcopyrightpermission[1]{\arxiv@copyrightfootnote{%
  \parindent\z@\parskip0.1\baselineskip
  This is the authors' version of the work, accepted for publication in
  \emph{\@acmBooktitle}. The definitive Version of Record will be available at
  \@formatdoi{\@acmDOI}.\par
  \copyright\ \@copyrightyear\ Copyright held by the owner/author(s).
  This work is licensed under a Creative Commons Attribution 4.0 International License.}}
\makeatother
\AddToHookNext{shipout/foreground}{%
  \put(0.5\paperwidth,-0.45in){\makebox[0pt]{\parbox{\textwidth}{\centering\small\itshape
    To appear in the 32nd ACM International Conference on Architectural Support for\\
    Programming Languages and Operating Systems (ASPLOS 2027)}}}}

\begin{document}

\title[Beyond LLM Serving: Characterizing VLA Workloads]{Beyond LLM Serving: Characterizing Vision-Language-Action Workloads for Embodied AI System Design}

\author{Seonghun Jung}
\orcid{0009-0004-2748-6589}
\affiliation{%
  \institution{KAIST}
  \city{Daejeon}
  \country{Republic of Korea}}
\email{shjung@casys.kaist.ac.kr}

\author{Sieun Moon}
\orcid{0009-0000-8257-4179}
\affiliation{%
  \institution{KAIST}
  \city{Daejeon}
  \country{Republic of Korea}}
\email{moon2121@kaist.ac.kr}

\author{Jiyoung Jeong}
\orcid{0009-0004-3794-6446}
\affiliation{%
  \institution{KAIST}
  \city{Daejeon}
  \country{Republic of Korea}}
\email{ji\_0.jeong@kaist.ac.kr}

\author{Jimin Lee}
\orcid{0009-0007-4478-0861}
\affiliation{%
  \institution{KAIST}
  \city{Daejeon}
  \country{Republic of Korea}}
\email{jimin-lee@kaist.ac.kr}

\author{Jaehyuk Huh}
\orcid{0000-0002-1742-047X}
\affiliation{%
  \institution{KAIST}
  \city{Daejeon}
  \country{Republic of Korea}}
\email{jhhuh@kaist.ac.kr}

\renewcommand{\shortauthors}{Seonghun Jung, Sieun Moon, Jiyoung Jeong, Jimin Lee, \& Jaehyuk Huh}

\begin{abstract}
Vision-language-action (VLA) models translate multimodal observations into low-level robot
actions. During robot operation, each control period sets an inference deadline, and overruns
leave the robot acting on stale observations, reducing task success. Meeting this deadline
motivates on-device or nearby edge execution, where a single robot requires batch-1 inference
outside the design point of LLM serving systems. Although VLA architectures combine familiar
vision-language, autoregressive, and diffusion-style components, their runtime behavior in
this batch-1 control setting remains uncharacterized. We characterize four representative VLA
models on an edge GPU server and two onboard SoCs, using single-inference profiling and
43,200 closed-loop episodes. Action tensor dimensionality determines whether a stage is
memory- or compute-bound, platform balance can shift that bottleneck, and GPU frequency
scaling yields a platform-dependent energy--latency sweet spot. In closed-loop operation,
overlapping inference with action execution creates an accuracy--speed--energy tradeoff, and
no configuration is Pareto-dominant across deployment SLOs. These results guide joint design
of VLA model architectures, hardware, and runtime policies.
\end{abstract}



\maketitle

\section{Introduction}

Embodied action generation models, often referred to as vision-language-action (VLA) models~\cite{bjorck2025gr00t, black2024pi_0, shukor2025smolvla, cheang2025gr, deng2025graspvla, bu2024towards, li2024cogact, wen2024diffusion, kim2025fine, zheng2025universal, song2025pd, li2025cogvla, kim2026cosmos}, translate multimodal observations into low-level robot control signals for manipulation tasks. Real-world manipulation requires smooth and continuous motion, so models designed for real robots increasingly produce not a single next command but an \emph{action chunk}, a fixed-length sequence of future commands that together form a short trajectory segment~\cite{zhao2023learning, jing2025mixture}. To generate such chunks, these models typically combine a large perception backbone with a specialized \emph{action-generation head}, a neural network that converts the backbone's latent representation into the action chunk. Recent architectures generate action chunks through iterative denoising over compact action tensors~\cite{bjorck2025gr00t, black2024pi_0, shukor2025smolvla, cheang2025gr, deng2025graspvla, bu2024towards, li2024cogact, wen2024diffusion}, single-pass parallel decoding~\cite{kim2025fine, song2025pd, li2025cogvla}, or unified diffusion over spatiotemporal video latents~\cite{kim2026cosmos}. These action-generation choices produce different latency and resource profiles that shape deployment decisions.

During deployment, these models operate in a closed loop: the robot continuously senses the environment, invokes the
model, executes the resulting commands, and updates its next input based on the physical
outcome~\cite{sapkota2025vision, ma2024survey}. The control period, roughly 33\,ms for
contact-rich manipulation, therefore acts as a recurring deadline rather than merely a per-request
latency target: inference that falls behind it leaves the robot acting on stale observations,
and task success degrades. Meeting this deadline favors on-device hardware or nearby
consumer-grade edge servers because remote serving adds round-trip delay to the same latency
budget. In the single-robot setting studied here, inference therefore runs at batch size one.

\newpage
LLM serving research has produced a mature toolbox, including continuous batching, paged
KV caches, and prefill-decode disaggregation, designed around concurrent autoregressive
requests and token-level execution~\cite{kwon2023efficient, zheng2024batchllm}. These
assumptions hold only partially in this setting: batch-1 serving provides no request concurrency, the
action-chunk models we evaluate produce their chunks non-autoregressively, and the dominant
stage varies with the architecture rather than with an autoregressive phase. Standard LLM
serving optimizations therefore apply selectively (Section~\ref{sec:discussion}). VLA model
families draw on familiar vision-language, autoregressive, and diffusion-style components, but
how these components interact under batch-1 closed-loop serving has not been characterized.

{\sloppy 
To fill this gap, we conduct a system-level workload characterization of VLA serving. We study four representative models, namely GR00T N1.6~\cite{bjorck2025gr00t}, $\pi_{0.5}$~\cite{intelligence2025pi_05}, OpenVLA-OFT~\cite{kim2025fine}, and Cosmos-Policy~\cite{kim2026cosmos}, that span three action-generation paradigms: iterative denoising, parallel decoding, and video generation. We evaluate these models on three platforms that span over an order of magnitude in power budget: RTX 4090, Jetson Thor, and Jetson AGX Orin~\cite{nvidia_jetson_modules}. Our methodology combines four measurement techniques that target different levels of the system stack: stage-level latency decomposition, roofline-based bottleneck attribution, DVFS~\cite{pillai2001real, tian2025clone, wang2025using} sensitivity analysis, and closed-loop deployment on MuJoCo~\cite{Todorov2002Mujoco} with LIBERO~\cite{liu2023libero} benchmarks, totaling 43,200 closed-loop episodes.
\par}

{\sloppy 
Our single-inference analysis shows that action tensor dimensions separate memory-bound
stages from compute-bound ones. Memory-bound stages tolerate lower GPU clocks.
Reducing memory bandwidth, however, raises latency and energy on both SoCs.
The same bottleneck differences shape throughput per joule in deployment.
Overlapping inference with action execution~\cite{shukor2025smolvla, tang2025vlash, black2025real} creates an accuracy--speed--energy tradeoff.
Accuracy degradation depends on model architecture, relative speedup is stable across
platforms, and energy scaling depends on platform characteristics.
\par}

To our knowledge, this is the first system-level workload characterization to connect
stage-level latency and roofline analysis with cross-platform DVFS sensitivity and closed-loop
deployment. Our contributions are as follows:
{\sloppy 

\noindent
\textbf{Cross-model characterization.} We show that VLA serving has no uniform bottleneck profile: \texttt{torch.compile} speedups differ by over 5× between action-head-dominated and backbone-dominated models.
This implies that compiler and kernel-level optimizations must be applied selectively per architecture rather than uniformly.

\noindent
\textbf{Microarchitectural analysis.} We show that the memory- versus compute-bound split is determined by action tensor dimensionality. Denoising-head GEMMs fall below the 163.9\,FLOP/B ridge point, whereas large-sequence backbone and video-generation DiT GEMMs exceed it. 
This provides a basis for accelerator bandwidth-to-compute provisioning in VLA-targeted hardware.

\noindent
\textbf{Platform and energy insights.} We show that GPU frequency scaling yields an energy sweet spot whose benefit depends on platform balance, as the same sweep saves 24--32\% energy on Orin but only 1--4\% on Thor, while memory-bandwidth throttling provides no energy benefit on either SoC. We further show that, when evaluated in isolation, assigning each inference phase its own GPU frequency can reduce both latency and energy when the phases exhibit different frequency sensitivities. However, in end-to-end execution, the overhead of switching GPU frequencies between phases can offset these gains.

\noindent
\textbf{Closed-loop tradeoff analysis.} We quantify a three-way tradeoff in which accuracy degrades by up to 56\,percentage points, speed improves by up to 3.9$\times$, and energy scales by up to 5.1$\times$ under increasing overlap, with no Pareto-dominant configuration across SLOs. This establishes that runtime systems must support SLO-aware operating-point selection rather than static configurations.

\par}

These observations motivate bottleneck-aware co-design across model architecture, hardware platform, and runtime policy for embodied AI system design.

\section{Background \& Motivation}
\label{sec:background}

\subsection{Embodied Action Generation Models}
\label{subsec:embodied_models}
Embodied action generation models produce low-level robot actions from visual observations and language instructions~\cite{sapkota2025vision, ma2024survey}.
Recent architectures first encode the multimodal observation and the instruction into an internal representation, and then generate actions from it. They differ in the architecture that builds this representation and in the action-generation interface.

In the VLM-based models, the \emph{perception backbone} typically combines a vision encoder (e.g., ViT~\cite{dosovitskiy2020image}, SigLIP~\cite{zhai2023sigmoid}) with a language model (e.g., Llama~2~\cite{touvron2023llama}, PaliGemma~\cite{beyer2024paligemma}) to produce contextual embeddings that capture both spatial and semantic information.
This stage performs a single forward pass through billions of parameters, inheriting the computational profile of large vision-language models.

The action-generation interface takes one of two forms: a sequence of discretized
\emph{action tokens} generated autoregressively, or an \emph{action chunk}, a fixed-length
sequence of future robot commands produced jointly in one inference~\cite{zhao2023learning, jing2025mixture}.

\textbf{Autoregressive action-token interface.}
Token-based VLAs such as RT-2~\cite{zitkovich2023rt} and vanilla OpenVLA~\cite{kim2024openvla}
represent a robot action as a sequence of discretized action tokens generated sequentially by
the backbone's language-model decoder. In the OpenVLA formulation we profile, the target
robot's action space has $D$ dimensions, and each dimension maps to one token. One action
therefore requires $D$ sequential token-generation steps: one prefill step followed by $D-1$
cached-decode passes.

This sequential decode path is difficult to amortize under high-rate control. On the
RTX\,4090, its six cached-decode passes take 140.90\,ms in our 7-dimensional setting. This is
3.2$\times$ the LLM prefill latency and accounts for 67\% of the 209.56\,ms inference
(Table~\ref{tab:openvla_ar}). Cached decode therefore dominates the latency of autoregressive
action generation.

The same decode path exhibits the familiar LLM resource split. Prefill is mostly
compute-bound, whereas cached decode is entirely memory-bound
(Section~\ref{subsec:roofline}). However, the fixed decode horizon leaves little opportunity
for KV-cache reuse. The cache grows by only $D-1$ positions within an action and is discarded
when the action completes.

Because this interface emits one $D$-token action per inference, producing $K$ future
actions repeats those $D$ sequential steps $K$ times. OpenVLA-OFT~\cite{kim2025fine} instead
emits an entire $K$-action chunk in one pass. Its reported action throughput on LIBERO is
roughly 26$\times$ that of autoregressive OpenVLA. This contrast motivates the action-chunk
interface, which amortizes one inference over $K$ future commands. Action-chunk models use
this interface but generate each chunk through one of three representative paradigms.

\begin{table}[t]
\centering
\caption{Latency decomposition and NCU roofline classification of autoregressive
OpenVLA on RTX\,4090 in eager mode.}
\vspace{-0.15in}
\label{tab:openvla_ar}
\small
\setlength{\tabcolsep}{2pt}
\begin{tabular}{@{} l r r r @{}}
\toprule
 & \textbf{Latency} & \multicolumn{2}{c}{\textbf{NCU GPU time (\%)}} \\
\cmidrule(lr){3-4}
\textbf{Stage} & \textbf{(ms)} & \makecell[r]{\textbf{Memory-}\\\textbf{bound}} &
\makecell[r]{\textbf{Compute-}\\\textbf{bound}} \\
\midrule
Vision encoder           &  16.03 &  23.9 & 76.1 \\
LLM prefill              &  43.50 &  13.6 & 86.4 \\
Cached decode (6 passes) & 140.90 & 100.0 &  0.0 \\
Other + preprocessing    &   9.13 &    -- &   -- \\
\midrule
E2E                      & 209.56 &    -- &   -- \\
\bottomrule
\end{tabular}
\vspace{-0.1in}
\end{table}

\textbf{Iterative denoising heads.}
This paradigm generates the action chunk through a multi-step
denoising process that invokes the same neural network $N$ times 
(typically 4--10) per action prediction.
Representative models refine a random initial tensor into the action trajectory,
using flow-matching objectives~\cite{lipman2023flow} (GR00T~\cite{bjorck2025gr00t},
$\pi_{0}$~\cite{black2024pi_0}, SmolVLA~\cite{shukor2025smolvla},
GR-3~\cite{cheang2025gr}, GraspVLA~\cite{deng2025graspvla}) or denoising
diffusion~\cite{NEURIPS2020_4c5bcfec, chi2025diffusion} (RoboDual~\cite{bu2024towards},
CogACT~\cite{li2024cogact}, DiffusionVLA~\cite{wen2024diffusion}).
Regardless of the training objective, the computational signature is 
shared: the state being denoised is a compact action trajectory tensor 
of shape $[\text{B},\;\text{action\_horizon},\;\text{action\_dim}]$ 
(batch size~1 in single-robot deployment), so each denoising step runs 
the network over this short, low-dimensional sequence.

\begin{figure}[t]
    \centering
    \includegraphics[trim={1mm 1mm 1mm 1mm}, clip, width=\columnwidth]{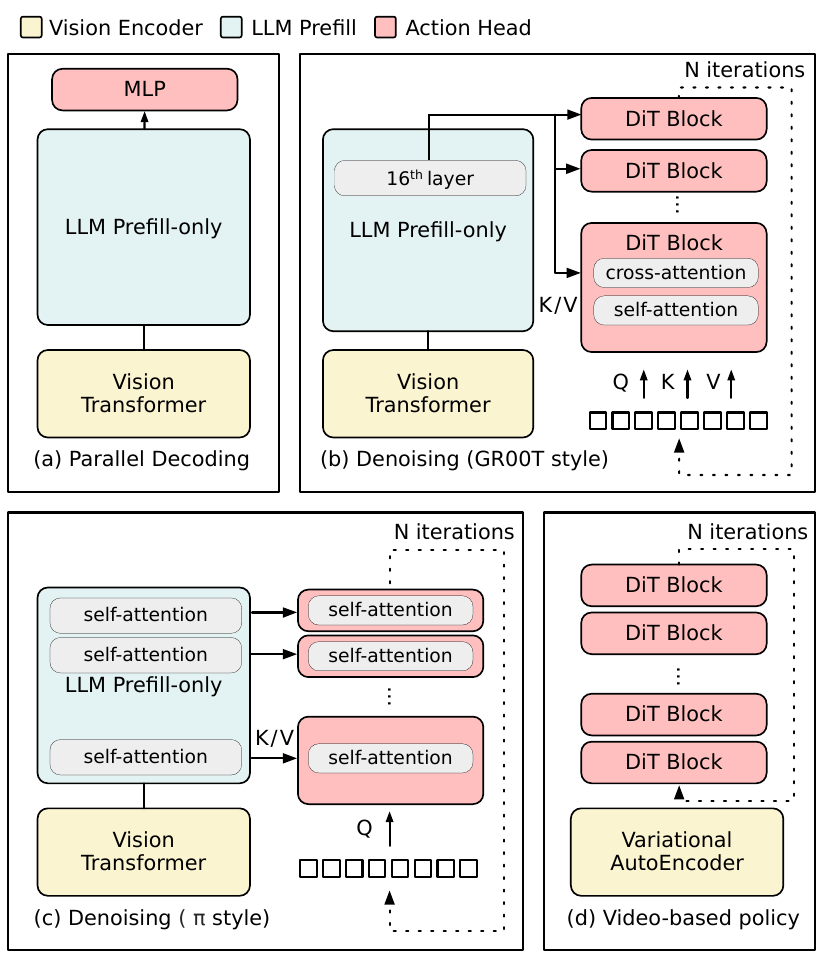}
    \begin{flushleft} %
    {\footnotesize\textsuperscript{$\dagger$} DiT: Diffusion Transformer}
    \end{flushleft}
    \vspace{-0.15in}
    \caption{Architectures and profiling decomposition of the four action-chunk models evaluated in closed loop.}
    \Description{Block diagrams compare parallel decoding, two iterative-denoising architectures, and a video-based policy, showing the backbone and action-generation stages used for profiling.}
    \label{fig:architecture_overview}
    \vspace{-0.05in}
\end{figure}

\textbf{Parallel decoding heads.}
This paradigm generates the entire action chunk in a single forward pass using parallel token prediction or regression.
Representative models include OpenVLA-OFT~\cite{kim2025fine} (L1 regression MLP), PD-VLA~\cite{song2025pd}, and CogVLA~\cite{li2025cogvla}.

\textbf{Unified video generation.}
This paradigm embeds robot actions in the latent sequence of a pretrained video diffusion
model, treating action generation as a joint video--action prediction problem within a shared
latent space~\cite{guo2024prediction, li2025unified}. Unlike the VLM-based models above,
Cosmos-Policy~\cite{kim2026cosmos} attaches no separate action head to a perception backbone:
it adapts Cosmos-Predict2~\cite{nvidia2025cosmosworldfoundationmodel} into a unified policy in which one generative model predicts
actions together with future visual observations.

Cosmos-Policy reuses the spatiotemporal VAE tokenizer of its pretrained video backbone to
encode image sequences into compact latent frames. It then injects robot proprioception,
action chunks, and future observations into dedicated slots of this latent sequence. A DiT
(Diffusion Transformer, a Transformer operating on latent image
patches)~\cite{peebles2023scalable} denoises the target slots conditioned on the
current-observation frames~\cite{kim2026cosmos}. Actions are then read directly from the
denoised action slots rather than decoded from generated RGB frames. The resulting tensors are far larger than the compact action trajectories of the
action-only denoising heads above: 9~temporal frames at $28{\times}28$ spatial resolution
yield $9{\times}28{\times}28 = 7{,}056$ token positions of 16 channels each.

For cross-model profiling, we separate VAE encoding from DiT denoising and group the
latter under the action-head stage, because the action-generation computation occurs within
the DiT. This is an operational decomposition for stage-level comparison: the DiT remains
part of Cosmos-Policy's unified video-diffusion architecture and also processes the other
latent slots.

Figure~\ref{fig:architecture_overview} illustrates the four action-chunk models
evaluated in closed loop: GR00T N1.6 (flow-matching DiT),
$\pi_{0.5}$~\cite{intelligence2025pi_05} (flow-matching Gemma Expert), OpenVLA-OFT (L1 regression MLP),
and Cosmos-Policy (video-generation DiT).
These four models span the three action-generation paradigms.
 
\subsection{Closed-Loop Policy Serving}
\label{subsec:closed_loop}
\begin{figure}[t]
    \centering
    \includegraphics[trim={1mm 1mm 1mm 1mm}, clip, width=\columnwidth]{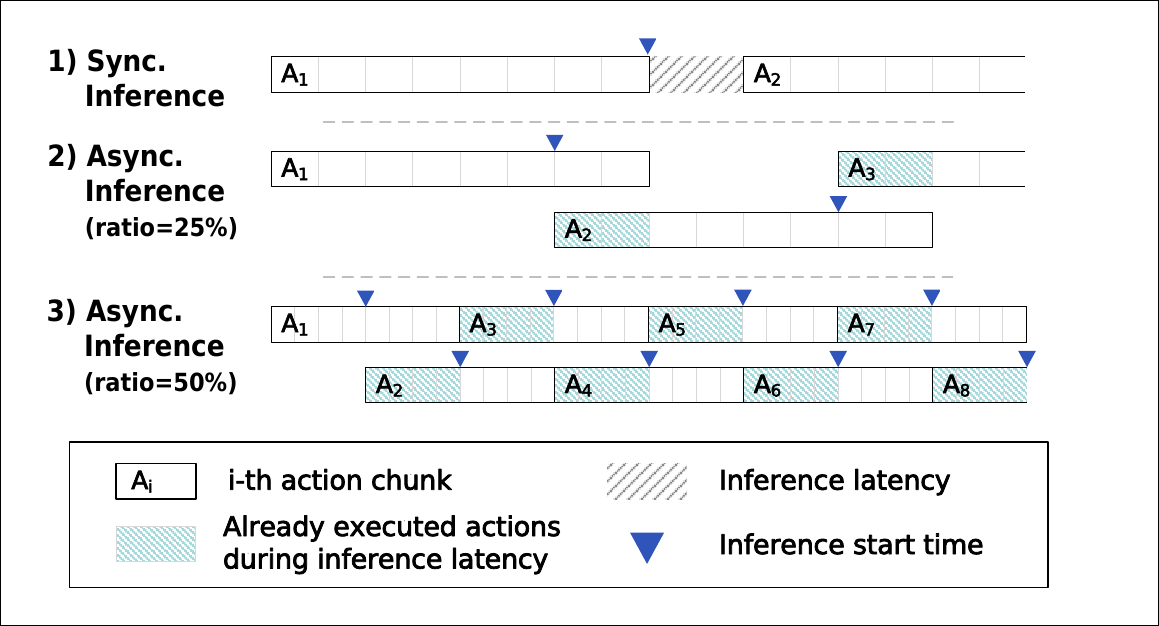}
    \vspace{-0.3in}
    \caption{Synchronous and asynchronous inference. In synchronous mode, the robot idles during inference (top). Asynchronous overlap initiates the next inference before the current chunk completes, with the overlap percentage governing when the next request is triggered (middle and bottom).}
    \Description{Timelines compare synchronous execution with asynchronous schedules that trigger the next inference after different fractions of the current action chunk have been consumed.}
    \label{fig:overlap_timeline}
    \vspace{-0.1in}
\end{figure}

When these policies drive a robot in closed loop, each inference must complete within a fixed control period, typically 33\,ms or less for the 30\,Hz rates required by contact-rich manipulation tasks.
Current policy models are far from meeting this budget~\cite{kim2025fine, wang2026real, zheng2026roboecc, yang2026dysl}: in our measurements, inference takes 117--396\,ms on an edge-server GPU (RTX\,4090) and 304--2603\,ms on an edge SoC (Jetson AGX Orin), yielding effective control frequencies of only 2.5--8.5\,Hz and 0.4--3.3\,Hz, respectively.

In a synchronous execution scheme, the control loop captures a sensor observation, invokes the policy model, and dispatches the resulting action chunk to the actuator before the next cycle begins~\cite{black2024pi_0, intelligence2025pi_05, bjorck2025gr00t, kim2025fine, team2024octo, zitkovich2023rt}.
This design maximizes observation freshness but, when inference overruns the control period, forces the pipeline to stall, repeating stale actions and introducing visible discontinuities in the robot's motion trajectory.
Unlike offline evaluation, where a simulator simply pauses during inference, real-world deployment offers no mechanism to stop time, so every millisecond of excess latency directly degrades control quality~\cite{black2025real}.

To bridge this gap, deployed systems employ \emph{asynchronous overlap}~\cite{shukor2025smolvla, tang2025vlash, black2025real}: the system issues the next inference request while the robot is still executing actions from the current chunk, hiding inference latency behind action execution (Figure~\ref{fig:overlap_timeline}).

However, asynchronous overlap introduces a tradeoff: because 
inference is initiated before the current chunk completes, the 
observation used for the next prediction ages while the robot 
continues executing~\cite{wang2026real}. We refer to this lag as 
\emph{observation staleness}, the interval between observation
capture and the execution of actions derived from it:
\begin{equation}
s = t_{\text{execute}} - t_{\text{capture}}.
\label{eq:staleness}
\end{equation}
More aggressive overlap enables faster control but forces the robot 
to act on increasingly stale observations, potentially degrading 
task accuracy~\cite{sendai2025leave}.
We refer to this as the \emph{accuracy--staleness tradeoff}.

\subsection{Motivation}

The workload described in Section~\ref{subsec:embodied_models} and Section~\ref{subsec:closed_loop} diverges from the assumptions that underlie LLM serving infrastructure in three ways. Each divergence motivates a distinct dimension of system-level characterization.

\noindent\textbf{M1. Action-generation heterogeneity requires stage-level bottleneck analysis.} LLM serving analyses decompose cost along the autoregressive phases of prefill and decode. That decomposition carries over to the action-token interface, which uses the same prefill--decode mechanism. The action-chunk models described in Section~\ref{subsec:embodied_models}, however, use heterogeneous iterative-denoising, parallel-decoding, and video-generation stages, which vary further within each paradigm through tensor shape and GEMM dimension choices. Characterizing them requires a stage-level decomposition organized around action-generation design.

\noindent\textbf{M2. Latency and task accuracy are coupled, not orthogonal.} In LLM serving, latency is a service-level property and response accuracy is a model property, and the two can be optimized independently. The accuracy--staleness tradeoff described in Section~\ref{subsec:closed_loop} removes this separation: per-inference latency affects the feasible overlap, which changes observation staleness and task success. Understanding this coupling requires measurement at the deployment level, not the per-inference level alone.

\noindent\textbf{M3. Deployment frequently spans heterogeneous platforms.} LLM serving has primarily targeted datacenter-class GPUs. Embodied action generation often runs on edge SoCs in addition to server GPUs, because autonomous operation may occur where network connectivity is unreliable or round-trip latency exceeds the control period. These platforms differ in compute, bandwidth, and power, so a characterization on a single tier does not readily transfer to others.

\section{Methodology}
\label{sec:methodology}
\subsection{Models and Benchmarks}
\label{subsec:models_benchmarks}
\begin{table}[t]
\centering
\caption{Architecture overview of evaluated models.}
\vspace{-0.15in}
\label{tab:models}
\small %
\setlength{\tabcolsep}{2pt} %

\begin{tabularx}{\columnwidth}{@{} l >{\raggedright\arraybackslash}X >{\raggedright\arraybackslash}X r @{}}
\toprule
\textbf{Model} & \textbf{Backbone} & \textbf{Action Head} & \textbf{Config\textsuperscript{1}} \\
\midrule
GR00T N1.6 & Cosmos-Reason VLM \newline {\footnotesize (Eagle ViT + LLM)} & DiT \newline {\footnotesize (Flow matching)} & 4/3B/16 \\ 
\addlinespace[3pt] %
$\pi_{0.5}$ & PaliGemma VLM \newline {\footnotesize (SigLIP+Gemma 2B)} & Gemma 300M \newline {\footnotesize (Flow matching)} & 10/3B/10 \\ 
\addlinespace[3pt]
OpenVLA-OFT & Prismatic VLM \newline {\footnotesize (SigLIP+DINOv2+\newline Llama~2)} & MLP \newline {\footnotesize (L1 regression)} & N/A/7B/8 \\
\addlinespace[3pt]
Cosmos-Policy & Wan2.1 VAE & DiT (Diffusion)\textsuperscript{2} & 5/2B/16 \\ 
\bottomrule
\end{tabularx}

\begin{minipage}{\columnwidth}
\footnotesize %
\textsuperscript{1}Config = Steps / $\sim$Params / Chunk. \textit{Steps}: denoising iterations; \textit{Chunk}: action steps per inference.\\
\textsuperscript{2}Cosmos-Policy employs a unified video diffusion architecture (Cosmos-Predict2-2B); VAE and DiT are listed separately for profiling clarity.
\end{minipage}
\vspace{-0.05in}
\end{table}

\begin{table}[t]
\centering
\caption{Hardware platforms and software stacks used in evaluation. }
\vspace{-0.15in}
\label{tab:hardware}
\small %
\setlength{\tabcolsep}{1.2pt}
\begin{tabular}{@{} >{\raggedright\arraybackslash}p{38pt} >{\raggedright\arraybackslash}p{73.5pt} >{\raggedright\arraybackslash}p{38.5pt} r >{\raggedright\arraybackslash}p{41.5pt} @{}}
\toprule
\textbf{Platform} & \textbf{Tier} & \textbf{Memory} & \textbf{Power} & \textbf{SW Stack} \\
\midrule
Jetson\newline AGX Orin & Power-efficient\newline SoC & 64\,GB LPDDR5   & 15--60  W& JetPack 6.2 \\ \addlinespace[2pt]
Jetson\newline Thor & High-performance\newline SoC & 128\,GB LPDDR5X & 70--130 W& JetPack 7.0  \\ \addlinespace[2pt]
RTX 4090        & Edge server   & 24\,GB GDDR6X   & 450     W& CUDA 13.1 \\
\bottomrule
\end{tabular}
\vspace{-0.05in}
\end{table}

Table~\ref{tab:models} lists the four embodied action generation models 
we evaluate. The set is chosen along two axes. First, it spans both 
action-generation mechanisms: iterative denoising (GR00T N1.6 and
$\pi_{0.5}$ over action tensors, Cosmos-Policy over video latents) and
single-pass parallel decoding (OpenVLA-OFT),
which impose different compute and control-loop profiles.
Second, it spans backbone families from standard 
vision-language models~\cite{touvron2023llama, li2025eagle, beyer2024paligemma} to video-generation 
models~\cite{nvidia2025cosmosworldfoundationmodel}, exposing how pretraining 
objective shapes deployment behavior. The goal is to cover diverse 
compute, memory, and control-loop behaviors rather than to maximize 
benchmark breadth.

{\sloppy
For the four models we evaluate end to end, we use the LIBERO-finetuned checkpoints from
each model's official repository rather than training our own variants.
Our measurements therefore reflect the deployment behavior of
community-adopted configurations.
We additionally profile the public \texttt{openvla-7b} checkpoint as an autoregressive
action-token baseline and exclude it from the closed-loop evaluation for its limited action
throughput (Section~\ref{subsec:embodied_models}).
\par}

We evaluate on four LIBERO~\cite{liu2023libero} suites (40 tasks in total),
which we organize into two task groups by task horizon for our analysis.
We refer to LIBERO-10 as the \emph{Long} group (10
long-horizon manipulation tasks with multi-step dynamic interactions), and we
pool LIBERO-Goal, LIBERO-Spatial, and LIBERO-Object into the \emph{Short}
group (30 short-horizon, shorter-episode tasks). We report the groups separately to compare
degradation across task horizons.
Tasks are the sampling units, and success rate (SR) is the mean of task-level
success rates. For latency, energy, and throughput, we first average successful episodes
per task and then average tasks.

\subsection{Platforms and Software}
\label{subsec:platforms}

Table~\ref{tab:hardware} covers an edge server and two onboard SoCs,
with power budgets ranging from 15\,W to 450\,W. We designate the RTX\,4090
as the primary platform for per-kernel roofline 
analysis~\cite{ding2019instruction, hill2019gables} because its discrete 
architecture exposes well-defined memory-bandwidth and 
compute-throughput ceilings, whereas the Jetson SoCs share DRAM 
between CPU and GPU and expose only aggregate bandwidth, complicating 
per-kernel attribution. Jetson Thor and AGX Orin~\cite{nvidia_jetson_modules} serve as deployment 
targets representing next- and current-generation mobile robots. All
platforms run PyTorch 2.10. The RTX\,4090 uses CUDA 13.1, while Thor 
and AGX Orin use JetPack 7.0 and 6.2 respectively, each bundling 
matched CUDA~\cite{cuda}, cuDNN~\cite{chetlur2014cudnn}, and TensorRT~\cite{nvidia_tensorrt} versions for its SoC generation.

\subsection{Measurement Methodology}
\label{subsec:measurement}
\begin{table}[t]
\centering
\caption{Measurement methodology summary. Each analysis targets a different level of the system stack; detailed configurations are deferred to the corresponding sections.}
\vspace{-0.15in}
\label{tab:measurement}
\newcolumntype{L}{>{\raggedright\arraybackslash}X} 

\begin{tabularx}{\columnwidth}{@{} l L L c @{}}
\toprule
\textbf{Analysis} & \textbf{Metric} & \textbf{Tool} & \textbf{Sec.} \\
\midrule
Latency decomp. & Wall-clock per stage (ms) & \texttt{perf\_counter} + \texttt{cuda.sync} & \ref{subsec:latency_decomp} \\ \addlinespace[1pt]
Roofline profiling & GPU kernel time, FLOPs, DRAM bytes & NVIDIA NCU (v2025.2.0) & \ref{subsec:roofline} \\ \addlinespace[1pt]
DVFS sensitivity & E2E latency per freq. point & \texttt{perf\_counter} + \texttt{sysfs} & \ref{subsec:soc_scaling} \\ \addlinespace[1pt]
Closed-loop & Success rate, measured energy & MuJoCo + NVML / \texttt{sysfs} & \ref{sec:closed_loop_inference} \\
\bottomrule
\end{tabularx}
\end{table}

\begin{table}[t]
\centering
\caption{Compilation configuration used for DVFS analysis and closed-loop evaluation. Compilation mode refers to \texttt{torch.compile} mode; ``eager'' denotes no compilation. Stage-level decomposition and roofline profiling use eager mode throughout.}
\vspace{-0.15in}
\label{tab:compile_config}
\small
\begin{tabularx}{\columnwidth}{@{} X l l @{}}
\toprule
\textbf{Model} & \textbf{Backbone} & \textbf{Action Head} \\
\midrule
GR00T N1.6     & Default           & Max-autotune \\
$\pi_{0.5}$    & Max-autotune      & Max-autotune \\
OpenVLA-OFT    & Max-autotune\textsuperscript{*} & N/A \\
Cosmos-Policy  & Eager             & Eager \\
\bottomrule
\addlinespace[1pt]
\multicolumn{3}{l}{\footnotesize \textsuperscript{*}Includes prewarm for variable-length task prompts.}
\end{tabularx}
\vspace{-0.05in}
\end{table}

Table~\ref{tab:measurement} summarizes measurement methodology by 
analysis, and Table~\ref{tab:compile_config} reports per-model 
compilation settings. All timing measurements insert GPU 
synchronization barriers (\texttt{cuda.synchronize}) at stage 
boundaries to attribute GPU work to the correct pipeline stage.
{\sloppy
Energy measurements cover the GPU on RTX\,4090. On each SoC, they cover the
compute-domain rails used by NVIDIA for module power capping:
\texttt{VDD\_GPU\_SOC} and \texttt{VDD\_CPU\_CV} on Orin~\cite{nvidia_orin_power}, and
\texttt{VDD\_GPU} and \texttt{VDD\_CPU\_SOC\_MSS} on Thor~\cite{nvidia_thor_power}.
These boundaries exclude total module input power. We therefore compare energy directly
within each platform and use cross-platform values only for relative trends.
\par}
For reproducibility, all models use each repository's official LIBERO-finetuned
checkpoint at its native input resolution, inference precision, and chunk length.
The per-phase frequency assignment in Section~\ref{subsubsec:phase_aware} assigns clocks
only to model-execution phases. Idle time is excluded because it is set by the control schedule
rather than by a single inference, and it is evaluated as part of the closed-loop timing
analysis in Section~\ref{sec:closed_loop_inference}.

{\sloppy
The per-model compilation settings in Table~\ref{tab:compile_config} 
reflect model-specific constraints that we document here for 
reproducibility. Cosmos-Policy runs in eager mode because compiling its 
TransformerEngine-based DiT slightly increases latency 
(Section~\ref{subsec:stage_compilation}) and its VAE tokenizer is not 
exposed as a compilable module. 
GR\-00T's backbone triggers tensor aliasing assertions under CUDA graph 
capture, so we apply \texttt{max-autotune} only to its action head and 
leave the backbone in default mode. OpenVLA-OFT compiles under 
\texttt{max-autotune}, but its task-description-dependent input length 
requires pre-warming all LIBERO task prompts before evaluation to 
cache each compilation variant. $\pi_{0.5}$ compiles cleanly under 
\texttt{max-autotune} across the full model.
\par}

These compiled configurations are used in analyses targeting 
deployment-realistic latency: DVFS sensitivity 
(Section~\ref{subsec:soc_scaling}) and closed-loop evaluation 
(Section~\ref{sec:closed_loop_inference}). Stage-level latency 
decomposition (Section~\ref{subsec:latency_decomp}) and roofline 
profiling (Section~\ref{subsec:roofline}) instead use eager mode, 
because compiler-induced kernel fusion and CUDA graph capture alter 
operator boundaries and obscure the per-kernel bottleneck structure 
those analyses target. Section~\ref{subsec:stage_compilation} 
explicitly compares both modes.

\section{Single-Inference Characterization}
\label{sec:single_inference}

{\sloppy
We characterize four action-chunk models through
stage-level latency decomposition (\S\ref{subsec:latency_decomp}), roofline bottlenecks
(\S\ref{subsec:roofline}), and edge-SoC DVFS sensitivity (\S\ref{subsec:soc_scaling}).
All measurements use random inputs matching each model's expected tensor shapes.
\par}

\subsection{Stage-Level Latency Decomposition of Action-Chunk Models}
\label{subsec:latency_decomp}
\begin{figure}[t]
    \centering
    \includegraphics[width=\columnwidth]{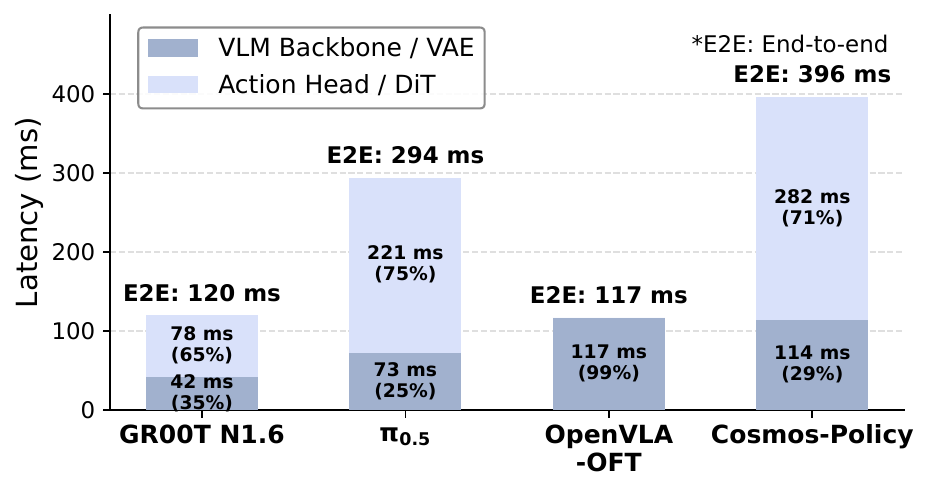}
    \vspace{-0.35in}
    \caption{End-to-end (E2E) latency breakdown in eager mode.}
    \Description{Stacked bars decompose eager-mode inference latency into model stages for the four action-chunk models.}
    \label{fig:eager_breakdown}
    \vspace{-0.05in}
\end{figure}

\subsubsection{Eager-mode breakdown}

{\sloppy
Figure~\ref{fig:eager_breakdown} decomposes the end-to-end inference latency of the
four action-chunk models into backbone and action-head contributions on the RTX\,4090.
Action generation accounts for 65\% of GR00T N1.6's latency, 75\% for $\pi_{0.5}$,
and 71\% for Cosmos-Policy.
OpenVLA-OFT is the exception: its single-pass MLP head completes in 0.6\,ms, making the 7\,B Llama~2 backbone the sole bottleneck (99\%).
\par}

\subsubsection{Per-stage compilation}
\label{subsec:stage_compilation}
\begin{figure}[t]
\centering
\includegraphics[width=\columnwidth]{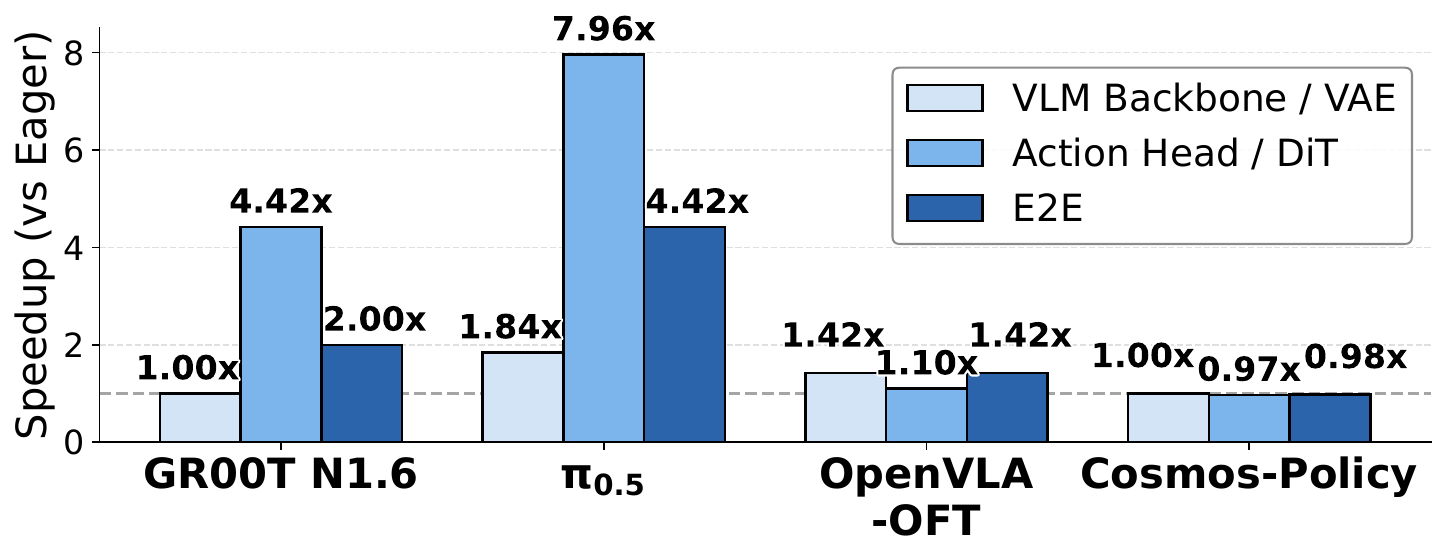}
\vspace{-0.35in}
\caption{Per-stage compilation speedup over eager execution by model.}
\Description{Grouped bars compare compilation speedup for the backbone and action-generation stages of each evaluated model.}
\label{fig:compile_per_stage}
\vspace{-0.05in}
\end{figure}

{\sloppy
Figure~\ref{fig:compile_per_stage} compares eager execution with independent compilation
of each eligible stage. Compiled stages use \texttt{max-autotune}, except GR00T N1.6's
backbone, which uses the default mode because of its CUDA-graph constraint
(Section~\ref{subsec:measurement}). Cosmos-Policy's VAE tokenizer is not exposed as a
compilable module, so it remains eager. Compilation accelerates the diffusion-based action heads most: 7.96$\times$ for $\pi_{0.5}$ and 4.42$\times$ for GR00T N1.6. 
OpenVLA-OFT instead gains 1.42$\times$ from backbone compilation. For Cosmos-Policy,
compiling the TransformerEngine-based DiT slightly increases latency (0.97$\times$).
The deployment configurations in Table~\ref{tab:compile_config} therefore retain
compilation only for stages where it improves latency.
\par}

\begin{takeawaybox}[takeaway:compile_bottleneck]
VLA pipeline architecture determines both the dominant latency stage and where
compilation helps. The iterative action head dominates and benefits most from compilation
in GR00T N1.6 and $\pi_{0.5}$; in OpenVLA-OFT, the backbone does. Compilation targets should
therefore be selected per architecture.
\end{takeawaybox}

\subsection{Compute--Memory Bottleneck Analysis of Action-Chunk Models}
\label{subsec:roofline}
Roofline analysis~\cite{williams2009roofline, ding2019instruction, hill2019gables}
classifies each GPU kernel as memory- or compute-bound from its arithmetic intensity
($\mathrm{AI} = \text{FLOPs}/\text{Bytes}$). The ridge point is the ratio of peak compute
throughput to peak memory bandwidth ($\pi_{\text{peak}}/\beta_{\text{peak}}$). Kernels whose
arithmetic intensity falls below the ridge point are memory-bound because memory bandwidth limits
performance before the compute units saturate. The RTX\,4090 provides 165.2\,TFLOP/s of FP16 throughput and 1.008\,TB/s of
memory bandwidth, giving a ridge point of 163.9\,FLOP/B.

\begin{figure}[t]
\centering
\includegraphics[width=\columnwidth]{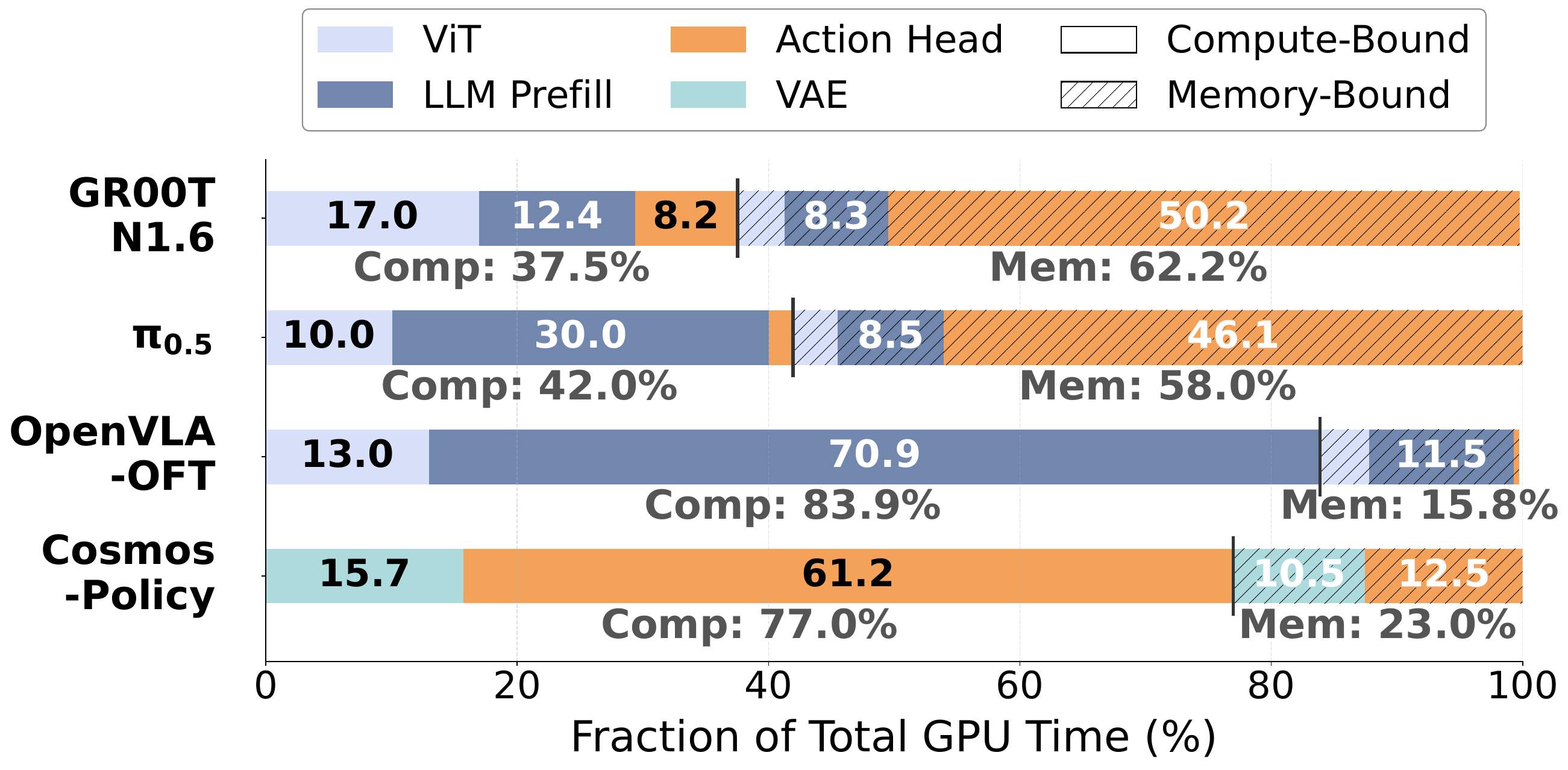}
\vspace{-0.25in}
\caption{Roofline classification by stage on RTX\,4090. Mem = memory-bound fraction of GPU time; Comp = compute-bound fraction.}
\Description{Stacked bars divide each model stage's GPU time into memory-bound and compute-bound fractions.}
\label{fig:roofline_bound_type}
\vspace{-0.05in}
\end{figure}

Table~\ref{tab:openvla_ar} applies the same classification to the autoregressive baseline.
Figure~\ref{fig:roofline_bound_type}
classifies GPU time by bottleneck type for each pipeline stage of the four action-chunk
models.
GR00T N1.6 and $\pi_{0.5}$ are predominantly memory-bound
(62\% and 58\% of GPU time), with the memory-bound fraction
concentrated in their action heads.
OpenVLA-OFT and Cosmos-Policy are predominantly compute-bound
(84\% and 77\%), dominated by the Llama~2 backbone and
video-generation DiT, respectively.

\begin{figure}[t]
\centering
\includegraphics[width=\columnwidth]{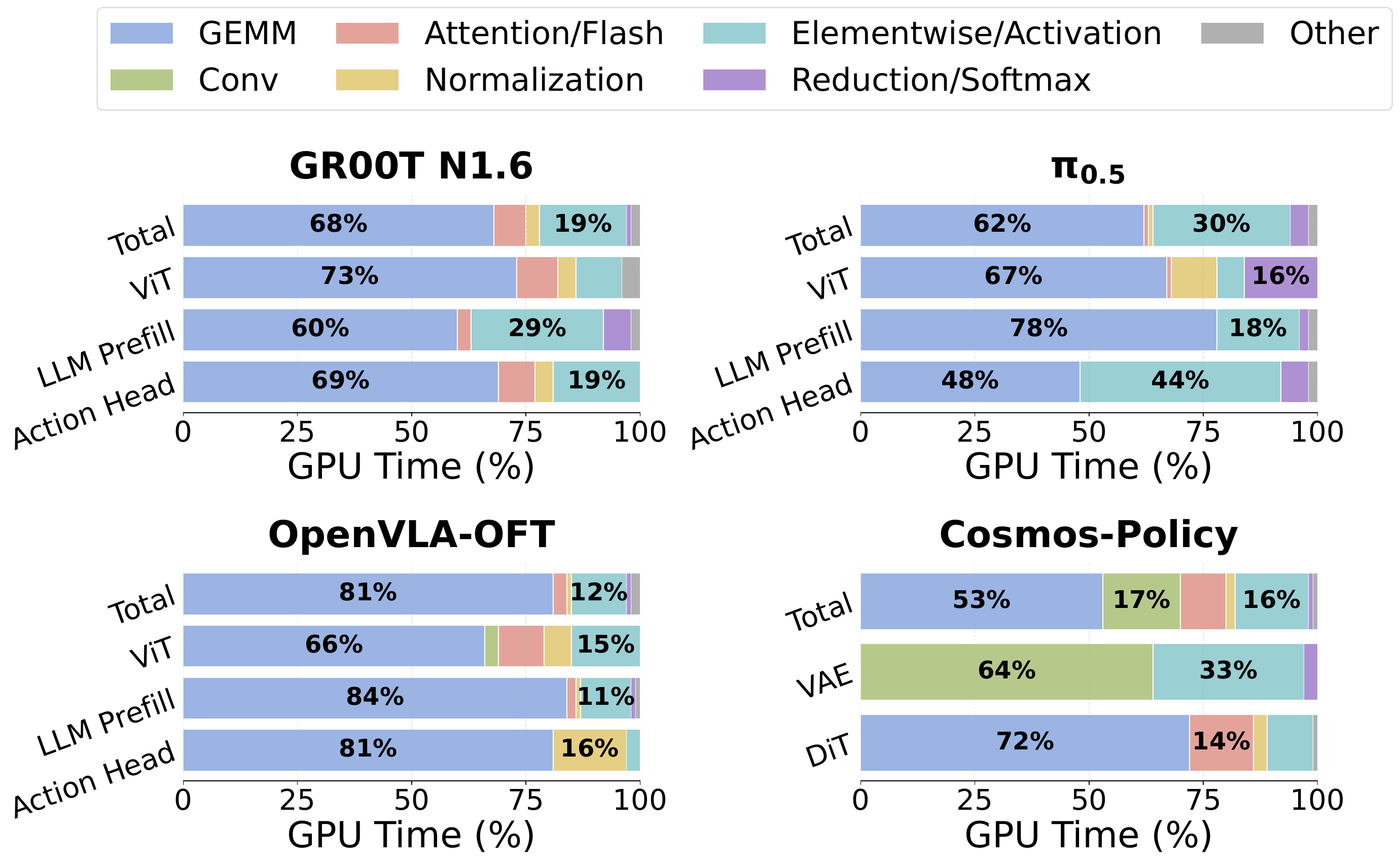}
\vspace{-0.25in}
\caption{Operator breakdown by stage and model on the RTX\,4090.}
\Description{Stacked bars show the fraction of GPU time spent in GEMM and other operator classes for each model stage.}
\label{fig:op_breakdown}
\vspace{-0.05in}
\end{figure}

GEMMs dominate GPU time in every model and stage
(Figure~\ref{fig:op_breakdown}), yet stages differ in the resource that limits them. Individual GEMM size therefore
determines whether a stage is memory- or compute-bound.

\begin{figure*}[t] 
\centering
\includegraphics[width=\textwidth]{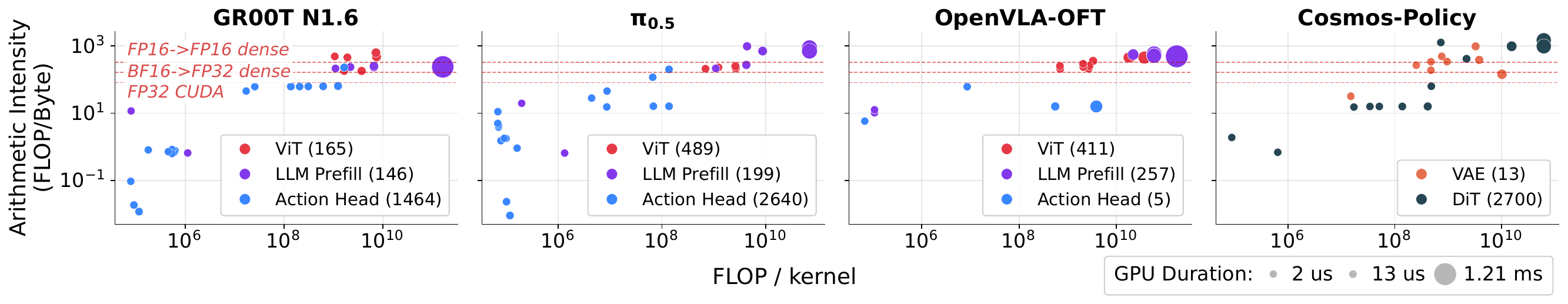}
\vspace{-0.3in}
\caption{GEMM size and arithmetic intensity on the RTX\,4090 roofline.}
\Description{A roofline scatter plot places GEMM kernels by operation count and arithmetic intensity, grouped by model and pipeline stage.}
\label{fig:gemm_scatter}
\vspace{-0.05in}
\end{figure*}

Figure~\ref{fig:gemm_scatter} plots per-GEMM arithmetic intensity 
against operation size. GR00T N1.6 and $\pi_{0.5}$ action-head GEMMs 
concentrate below the ridge point, while OpenVLA-OFT backbone and 
Cosmos-Policy DiT GEMMs cluster above it.
The intensity gap is determined by the tensor dimensions processed by each stage.
GR00T N1.6's DiT processes a short sequence of action and state tokens.
At batch size one, its linear layers reuse their weights across only a few tokens,
limiting arithmetic intensity. The denoising loop repeats these low-intensity
operations at each step. $\pi_{0.5}$ similarly processes a short suffix sequence
at each step, and its action-head GEMMs also fall below the ridge point.
OpenVLA-OFT processes all tokens in full self-attention
through a 7\,B backbone, and Cosmos-Policy denoises a
spatiotemporal latent with 7{,}056 effective positions. Both
produce GEMMs above the ridge point.

\begin{takeawaybox}[takeaway:mem_compute_bound]
The memory/compute-bound distinction arises from tensor dimensionality:
iterative denoising over compact action tensors is bottlenecked by memory bandwidth,
while large-sequence processing saturates compute units.
\end{takeawaybox}

\subsection{Cross-Platform Resource Sensitivity on Edge SoCs}
\label{subsec:soc_scaling}
\begin{figure}[t]
\centering
\includegraphics[width=\columnwidth]{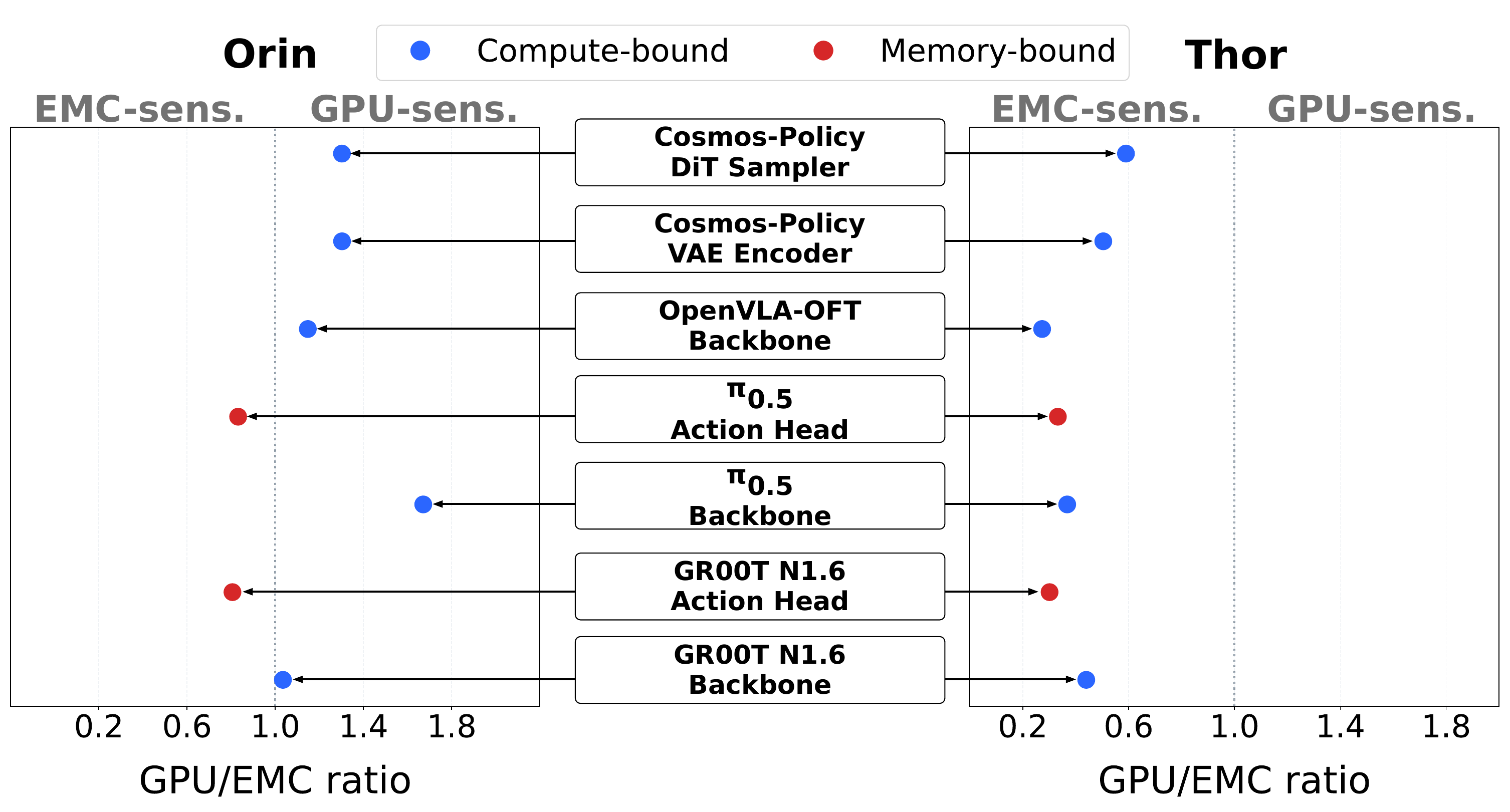}
\vspace{-0.2in}
\caption{Per-stage GPU/EMC sensitivity ratio $R$
    (Eq.~\ref{eq:ratio}) on Orin and Thor.
    Points are colored by the roofline-predicted bottleneck
    (memory- or compute-bound) from RTX\,4090 eager-mode profiling.}
\Description{Points compare per-stage sensitivity to GPU and memory frequency on Orin and Thor, colored by the RTX 4090 roofline classification.}
\label{fig:soc_scaling}
\vspace{-0.05in}
\end{figure}

The roofline analysis above was conducted on a discrete GPU
(RTX\,4090) with separate compute and memory subsystems.
Edge SoCs share a unified memory hierarchy and expose three
independently clocked domains, each scalable via
DVFS~\cite{pillai2001real, tian2025clone, wang2025using}: CPU cores, GPU, and the EMC
(external memory controller, which governs DRAM
bandwidth)~\cite{nvidia_orin_power, nvidia_thor_power}.
This subsection examines whether the roofline-predicted bound
types manifest as measurable resource sensitivity on Jetson
Thor and AGX Orin.

\textbf{Definitions.}
We define the \emph{sensitivity} of a pipeline stage (or
end-to-end inference) to a clock domain as the normalized
latency when that domain is set to its minimum frequency, with
all other domains held at maximum:

\begin{equation}
  S_{\text{domain}} \;=\; \frac{T_{\text{domain}=f_{\min}}}{T_{\text{all}=f_{\max}}}.
  \label{eq:sensitivity}
\end{equation}
A higher value indicates greater dependence on that resource.
To compare the two dominant hardware resources within a single
scalar, we define the \emph{GPU/EMC sensitivity ratio}:
\begin{equation}
  R \;=\; \frac{S_{\text{GPU}}}{S_{\text{EMC}}}.
  \label{eq:ratio}
\end{equation}
A ratio $R < 1$ indicates that the stage is more sensitive to
memory bandwidth (EMC) than to compute throughput (GPU), while $R > 1$
indicates GPU-compute dominance.

\textbf{Experimental setup.}
All measurements use compiled execution with the per-model
configurations listed in Table~\ref{tab:compile_config}.
GPU and EMC frequencies are swept independently while all other
domains remain at maximum.
The swept ranges are:
Orin GPU 1300\,$\rightarrow$\,306\,MHz (4.2$\times$),
Orin EMC 3199\,$\rightarrow$\,665.6\,MHz (4.8$\times$),
Thor GPU 1575\,$\rightarrow$\,432\,MHz (3.6$\times$), and
Thor EMC 4266\,$\rightarrow$\,665.6\,MHz (6.4$\times$).
Because the two platforms differ in frequency range, absolute
sensitivity values are not directly comparable across platforms,
so we use the ratio~$R$ to compare resource binding within and
across platforms.

\subsubsection{Per-Stage Sensitivity Ratio}
\label{subsubsec:stage_ratio}

{\sloppy 
Figure~\ref{fig:soc_scaling} plots the GPU/EMC sensitivity
ratio~$R$ for each pipeline stage on both platforms, with points
colored by the roofline-predicted bottleneck (memory- or
compute-bound) obtained from RTX\,4090 eager-mode profiling (Section~\ref{subsec:roofline}).

\par}

{\sloppy
On Orin, the sensitivity ratio separates memory-bound from
compute-bound stages around $R \approx 1$.
Memory-bound action heads
(GR00T N1.6 $R{=}0.81$, $\pi_{0.5}$ $R{=}0.83$) exhibit
stronger relative EMC sensitivity than all compute-bound
stages (backbones, Cosmos-Policy VAE and DiT), which range
from $R{=}1.03$ to $R{=}1.67$.
The roofline-predicted ordering is also observed in
DVFS-derived resource sensitivity on this platform.
\par}

On Thor, absolute $R$ values drop below~1 for all
stages, reflecting the platform's higher
compute-to-bandwidth ratio: GPU throttling from 1575 to
432\,MHz yields only 1.3--2.3$\times$ slowdown across a
3.6$\times$ frequency range, while EMC throttling to
665.6\,MHz causes 3.3--4.7$\times$ slowdown.
Despite this platform-wide shift, the relative
ordering among stages largely tracks the roofline
classification: stages with sub-ridge arithmetic intensity
(GR00T N1.6 action head $R{=}0.30$, $\pi_{0.5}$ action head
$R{=}0.33$) show lower~$R$ than most compute-bound stages
(GR00T N1.6 backbone $R{=}0.44$, Cosmos-Policy VAE
$R{=}0.50$, DiT $R{=}0.59$).
The ordering is not perfect. The OpenVLA-OFT backbone
($R{=}0.27$) falls below both memory-bound action heads.
The overall trend, however,
indicates that per-kernel arithmetic intensity continues to
modulate relative EMC sensitivity even when the platform
places all stages in an EMC-dominant regime.

Together, the Orin and Thor results show that platform balance shifts absolute
sensitivity, while arithmetic intensity still organizes the broad stage-level trend.
 
\begin{takeawaybox}[takeaway:platform_bottleneck_shift]
Whether a stage is limited by memory bandwidth or by compute throughput
depends on both its arithmetic intensity and the platform's
compute-to-bandwidth ratio. The same stage can therefore hit different
bottlenecks across platforms, while the relative ordering of stages by
intensity is largely preserved.
\end{takeawaybox}

\subsubsection{Energy--Latency Tradeoff under Frequency Scaling}
\label{subsubsec:energy_tradeoff}

\begin{figure}[t]
\centering
\includegraphics[width=\columnwidth]{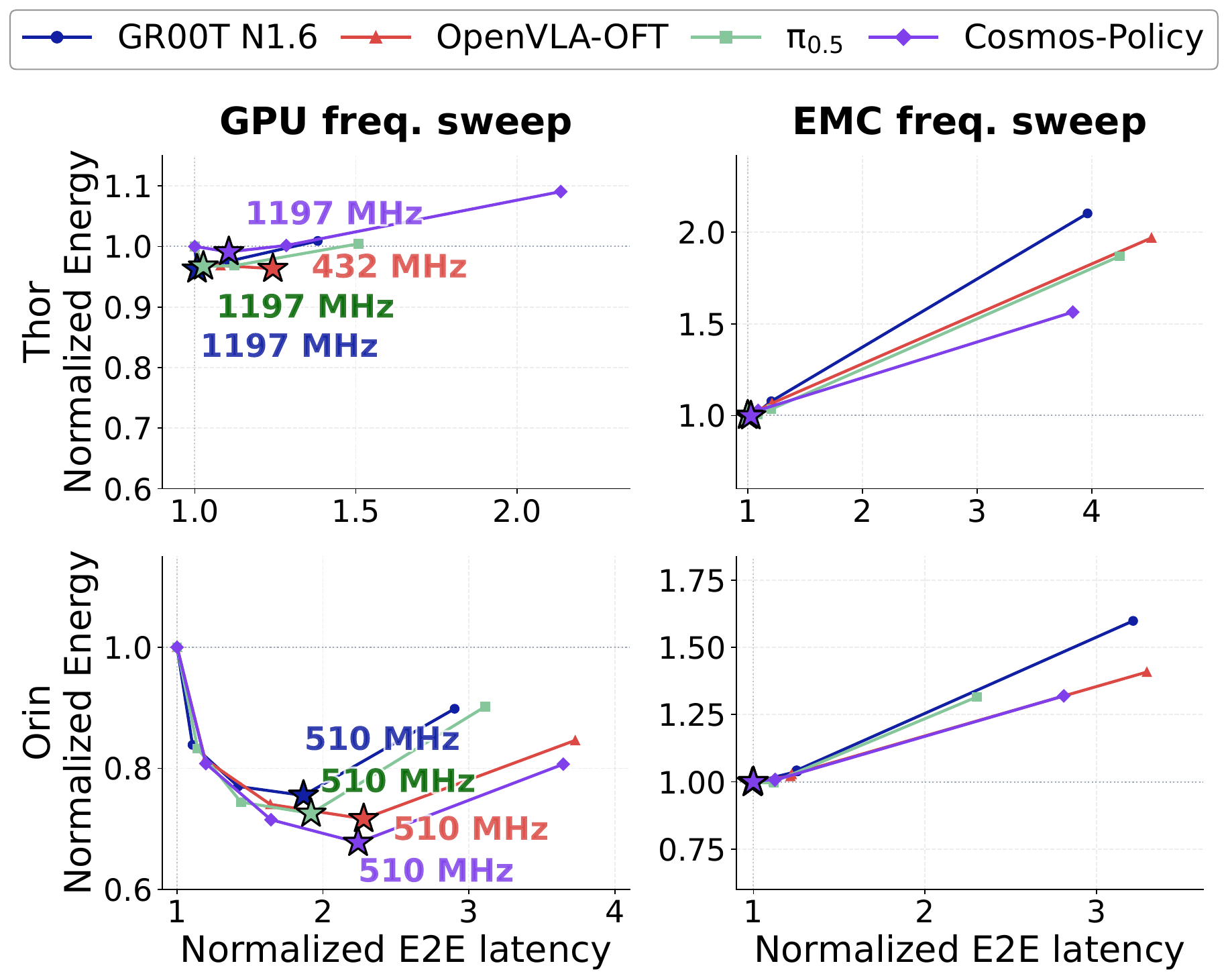}
\vspace{-0.2in}
\caption{Energy--latency tradeoffs under GPU and EMC frequency sweeps on Jetson AGX Orin and Jetson Thor.}
\Description{Energy-latency curves compare GPU-frequency and memory-frequency sweeps on the two edge SoCs.}
\label{fig:soc_energy_latency}
\vspace{-0.05in}
\end{figure}

Per-inference energy equals average power multiplied by
execution time.
Figure~\ref{fig:soc_energy_latency} plots normalized
per-inference energy against normalized end-to-end latency
for GPU and EMC frequency sweeps on both platforms.

\begin{figure*}[t]
\centering
\includegraphics[width=\textwidth]{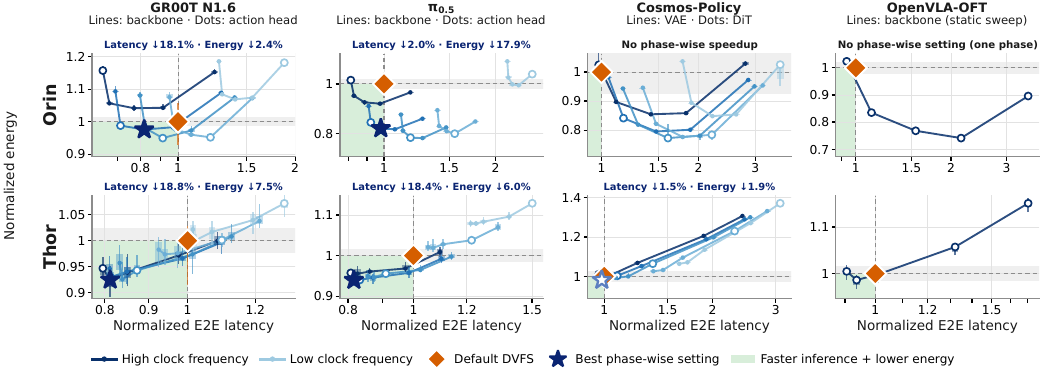}
\vspace{-0.25in}
\caption{Transition-free latency--energy tradeoffs under the three-rule phase-aware
policy on Orin and Thor, normalized to the default platform DVFS configuration (lower is better). Open stars
denote energy changes within measurement resolution, and candlesticks show energy
variability.}
\Description{Eight panels plot normalized latency against normalized energy for phase-aware GPU-frequency assignments on Orin and Thor across four models.}
\label{fig:phase_aware_grid}
\vspace{-0.05in}
\end{figure*}

\textbf{GPU throttling traces a convex energy--latency curve.}
On Orin, all four models reach minimum energy at 510\,MHz,
saving 24--32\% of per-inference energy at 1.9--2.2$\times$
latency cost.
The minimum exists because power drops faster than latency
rises while memory-bound stages are less sensitive to GPU clock
reduction. Below 510\,MHz, latency growth in compute-bound
stages overtakes the power saving.
On Thor, the curve is much flatter: the energy-optimal point
(1197\,MHz for three models, 432\,MHz for OpenVLA-OFT) yields
only 1--4\% savings, consistent with the observation that GPU
throttling has limited impact on latency across all stages on
this platform.
 
\textbf{EMC throttling offers no energy benefit.}
Energy rises monotonically on both platforms as EMC frequency
decreases, reaching +32\% to +60\% on Orin and +56\% to +110\%
on Thor at minimum frequency.
Bandwidth starvation inflates latency across all stages, and
this increase outweighs any power reduction from lower DRAM
clock.

\begin{takeawaybox}[takeaway:dvfs_gpu_scaling]
Among the evaluated DVFS controls, only GPU frequency scaling yields an
energy--latency sweet spot. The attainable energy savings depend on the platform's
compute-to-bandwidth balance: 24--32\% on Orin versus 1--4\% on Thor.
\end{takeawaybox}

\subsubsection{Phase-Aware Frequency Policy}
\label{subsubsec:phase_aware}
\begin{table}[t]
\centering
\caption{Measured GPU-frequency transition latency and its share of baseline E2E latency
for two transitions per inference. OpenVLA-OFT has no intra-inference transition and is
omitted.}
\vspace{-0.15in}
\label{tab:transition_overhead}
\small
\setlength{\tabcolsep}{4pt}
\begin{tabular}{@{} l r r c c c @{}}
\toprule
& \multicolumn{2}{c}{\textbf{Transition latency}} & \multicolumn{3}{c}{\textbf{Share of E2E latency}} \\
\cmidrule(lr){2-3} \cmidrule(l){4-6}
\textbf{Platform} & \makecell[r]{Per\\transition} & \makecell[r]{Per\\inference}
& \makecell{GR00T\\N1.6} & \makecell{$\pi_{0.5}$\\~} & \makecell{Cosmos-\\Policy} \\
\midrule
Orin & 3.5\,ms & 7.0\,ms & 2.5\% & 1.2\% & 0.3\% \\
Thor & 8.6\,ms & 18.0\,ms & 13.6\% & 7.8\% & 1.7\% \\
\bottomrule
\end{tabular}
\vspace{-0.05in}
\end{table}

{\sloppy
Sections~\ref{subsubsec:stage_ratio} and~\ref{subsubsec:energy_tradeoff} show that the phases of
one inference place different demands on host dispatch, memory bandwidth, and GPU compute. We derive
three rules from these measurements and evaluate them together. \emph{(i) CPU at maximum
frequency.} Iterative action heads repeatedly dispatch small kernels and gain most from
compilation and kernel fusion (Takeaway~\ref{takeaway:compile_bottleneck}), so keeping the CPU at
maximum frequency reduces host-side dispatch delay. \emph{(ii) EMC at maximum frequency.} EMC
throttling increases per-inference energy on both SoCs
(Section~\ref{subsubsec:energy_tradeoff}), so we keep EMC at maximum frequency. \emph{(iii) GPU
frequency selected for each phase.} Phases with different resource demands
respond differently to GPU clock (Takeaway~\ref{takeaway:platform_bottleneck_shift}), so we
allow each phase to use a separate clock. Together, the fixed CPU and EMC settings and phase-specific GPU
settings form the phase-aware policy evaluated below.

\par}

\textbf{Sweep and selection.}
With the first two rules active, we sweep five candidate GPU clocks over the range used by each
platform's governor: 306--1300\,MHz on Orin and 315--1575\,MHz on Thor. A two-phase model
assigns one of the five clocks to each phase independently, so we evaluate all 25 clock
combinations. The baseline is the default platform DVFS configuration, which leaves the GPU, EMC, and
CPU frequencies under their default governors and enables CPU idle. We select the minimum-energy
phase-wise assignment whose latency is lower than this baseline. Assignments that give both
phases the same clock reduce to static whole-inference scaling, so we exclude them from
selection. Figure~\ref{fig:phase_aware_grid}
reports the sweep and the selected points.

\textbf{Transition-free evaluation.}
The reported latency excludes measured clock-switch time. Phase-aware energy is recombined as
$\sum_i t_i P_i$ from per-phase power measured at each static clock because the \mbox{INA3221}
cannot resolve the ${\sim}$3.5\,ms (Orin) and ${\sim}$8.6\,ms (Thor) transitions well enough to attribute their energy between phases.
Figure~\ref{fig:phase_aware_grid} therefore measures the operating-point opportunity without
transition cost. Section~\ref{subsubsec:transition_overhead} quantifies the measured transition
latency.

\textbf{Models with dissimilar phases offer favorable operating points under the phase-aware policy.}
{\sloppy GR00T\,N1.6 and $\pi_{0.5}$ combine a compute-heavy backbone with an iteration-heavy
action head whose resource demands differ from those of the backbone.
Figure~\ref{fig:phase_aware_grid} shows that, for both models on both SoCs, phase-wise
assignments under the phase-aware policy reduce latency and energy relative to default DVFS
in the transition-free evaluation. On Orin, $\pi_{0.5}$ reduces energy by 17.9\% and latency
by 2.0\%. On Thor, GR00T N1.6 reduces energy by 7.5\% and latency by 18.8\%, while $\pi_{0.5}$
reduces energy by 6.0\% and latency by 18.4\%.
\par}

\textbf{Switching needs distinct, long phases.}
Cosmos-Policy's VAE and DiT are both dominated by large-GEMM work. On Orin, no phase-wise
assignment is faster than the baseline because lowering either clock increases latency. On Thor,
the selected point is 1.5\% faster and 1.9\% lower energy, but the energy difference lies within
the 2.2\% measurement resolution for that configuration. OpenVLA-OFT presents a different limit: its action head
is a single 0.6\,ms MLP pass (Section~\ref{subsec:latency_decomp}), shorter than one frequency
transition, so only the backbone is long enough for GPU-clock assignment.

\subsubsection{Transition Overhead}
\label{subsubsec:transition_overhead}

{\sloppy Table~\ref{tab:transition_overhead} reports median transition overheads of
3.5\,ms on Orin and 8.6\,ms on Thor, and the resulting E2E share reaches 13.6\% on Thor.
We measure transition latency but do not separately attribute transition energy.}

\textbf{Transition latency limits gains from short phases.}
On Thor, GR00T\,N1.6 spends 13.6\% of baseline E2E latency on two transitions, which consumes most
of its 18.8\% transition-free latency gain. Cosmos-Policy's longer inference spends less than
2\% on the same two transitions. Phase heterogeneity determines the available operating-point
gain, while phase duration relative to transition latency limits the realizable portion. We report this margin
without implementing a runtime policy. Both axes of Figure~\ref{fig:phase_aware_grid} remain transition-free upper bounds.

The analysis thus far has treated each inference in isolation.
In closed-loop deployment, per-inference latency governs overlap aggressiveness, shaping both control frequency and observation staleness.
Section~\ref{sec:closed_loop_inference} quantifies this interaction.

\begin{takeawaybox}[takeaway:phase_aware]
Phase-aware frequency control is useful when GPU phases differ in clock sensitivity
and are long enough to amortize frequency transitions. GR00T N1.6 and $\pi_{0.5}$ meet these
conditions, saving 2.4--17.9\% energy while running faster when the clock-switch cost is
excluded. The switching time itself is platform-dependent and can consume much of this
latency gain when phases are short.
\end{takeawaybox}

\begin{table}[t]
\centering
\caption{Worst-case round-trip time ($\text{RTT}_{\max}$) and control periods per overlap level. $\text{RTT}_{\max}$ averages each episode's maximum RTT under synchronous operation and covers observation transfer, server-side pre- and post-processing, and compiled inference. Control periods are derived via Eq.~\ref{eq:control_period}.}
\vspace{-0.15in}
\label{tab:control_period}
\small %

\newcolumntype{R}{>{\raggedleft\arraybackslash}p{54pt}}
\newcolumntype{D}{>{\raggedleft\arraybackslash}p{28pt}}
\setlength{\tabcolsep}{1pt} %

\begin{tabularx}{\columnwidth}{@{} l l R D D D @{}}
\toprule
& & & \multicolumn{3}{c}{\textbf{Control Period (ms)}} \\
\cmidrule(l){4-6}
\textbf{Model} & \textbf{Platform} & \textbf{$\text{RTT}_{\max}$ (ms)} & \textbf{10\%} & \textbf{25\%} & \textbf{50\%} \\
\midrule
GR00T N1.6    & RTX 4090 & 142.8 & 107.1 & 53.5 & 26.8 \\
              & Thor     & 126.7 & 95.1 & 47.5 & 23.8 \\
              & Orin     & 206.8 & 155.1 & 77.6 & 38.8 \\
\addlinespace[2pt]
$\pi_{0.5}$   & RTX 4090 & 136.1 & 204.1 & 68.0 & 40.8 \\
              & Thor     & 180.6 & 270.8 & 90.3 & 54.2 \\
              & Orin     & 457.3 & 686.0 & 228.7 & 137.2 \\
\addlinespace[2pt]
OpenVLA-OFT   & RTX 4090 & 148.7 & 223.0 & 111.5 & 55.8 \\
              & Thor     & 306.8 & 460.1 & 230.1 & 115.0 \\
              & Orin     & 528.2 & 792.2 & 396.1 & 198.1 \\
\addlinespace[2pt]
Cosmos-Policy & RTX 4090 & 442.9 & 332.2 & 166.1 & 83.0 \\
              & Thor     & 1023.0 & 767.2 & 383.6 & 191.8 \\
              & Orin     & 2558.4 & 1918.8 & 959.4 & 479.7 \\
\bottomrule
\end{tabularx}
\vspace{-0.1in}
\end{table}

\section{Closed-Loop Serving Tradeoffs}
\label{sec:closed_loop_inference}

This section maps single-inference costs to closed-loop behavior. We study
asynchronous overlap (\S\ref{subsec:overlap_design}), accuracy--speed--energy
tradeoffs (\S\ref{subsec:tradeoff}), throughput--energy efficiency
(\S\ref{subsec:throughput_energy}), and SLO-constrained operating points
(\S\ref{subsec:slo}).

\begin{figure}[t]
\centering
\includegraphics[width=\columnwidth]{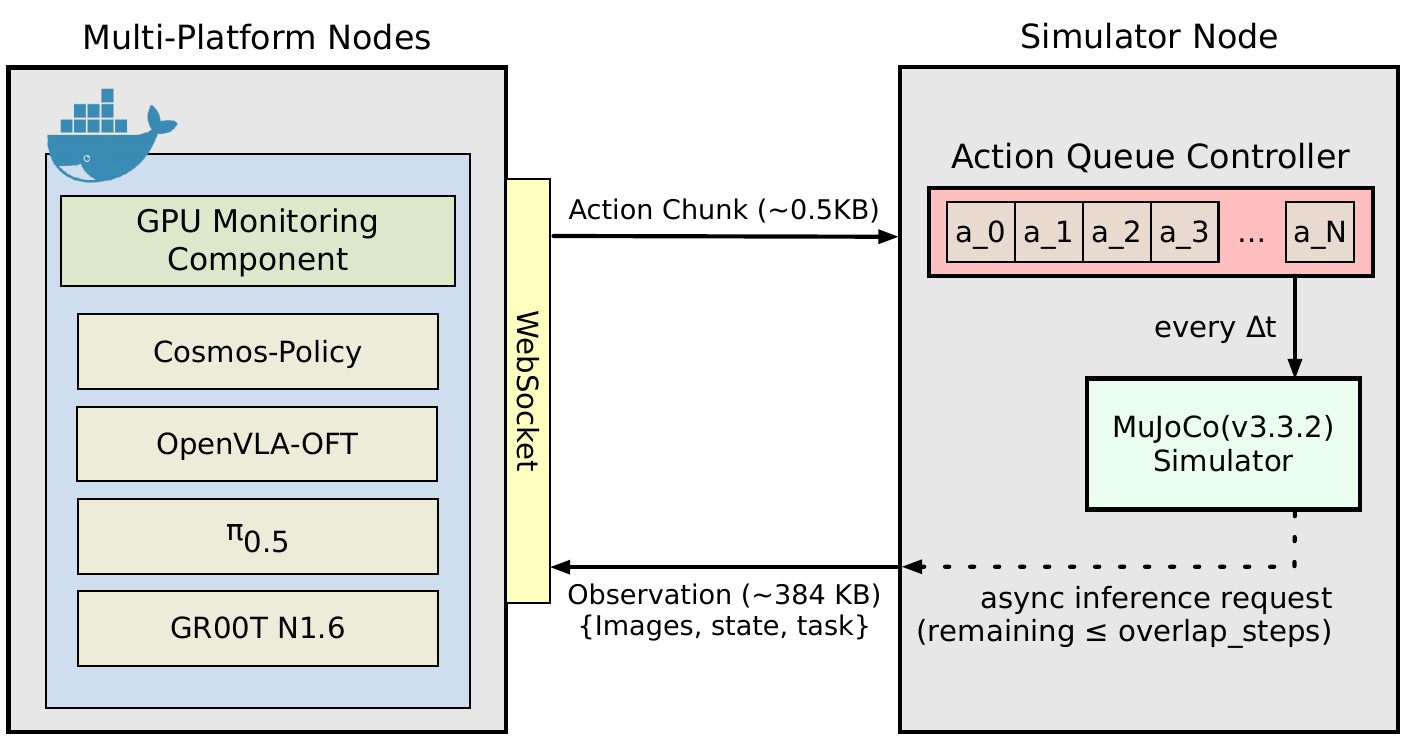}
\vspace{-0.2in}
\caption{System architecture of the closed-loop benchmark.
The client node runs a MuJoCo simulator and consumes actions
from a FIFO queue at fixed intervals~$\Delta t$, while
asynchronously prefetching new action chunks from
multi-platform model servers over WebSocket.}
\Description{A client-server diagram shows a MuJoCo client consuming actions from a FIFO queue while asynchronously requesting action chunks from model servers.}
\label{fig:system_architecture}
\vspace{-0.05in}
\end{figure}

\begin{figure*}[t]
\centering
\includegraphics[width=\textwidth]{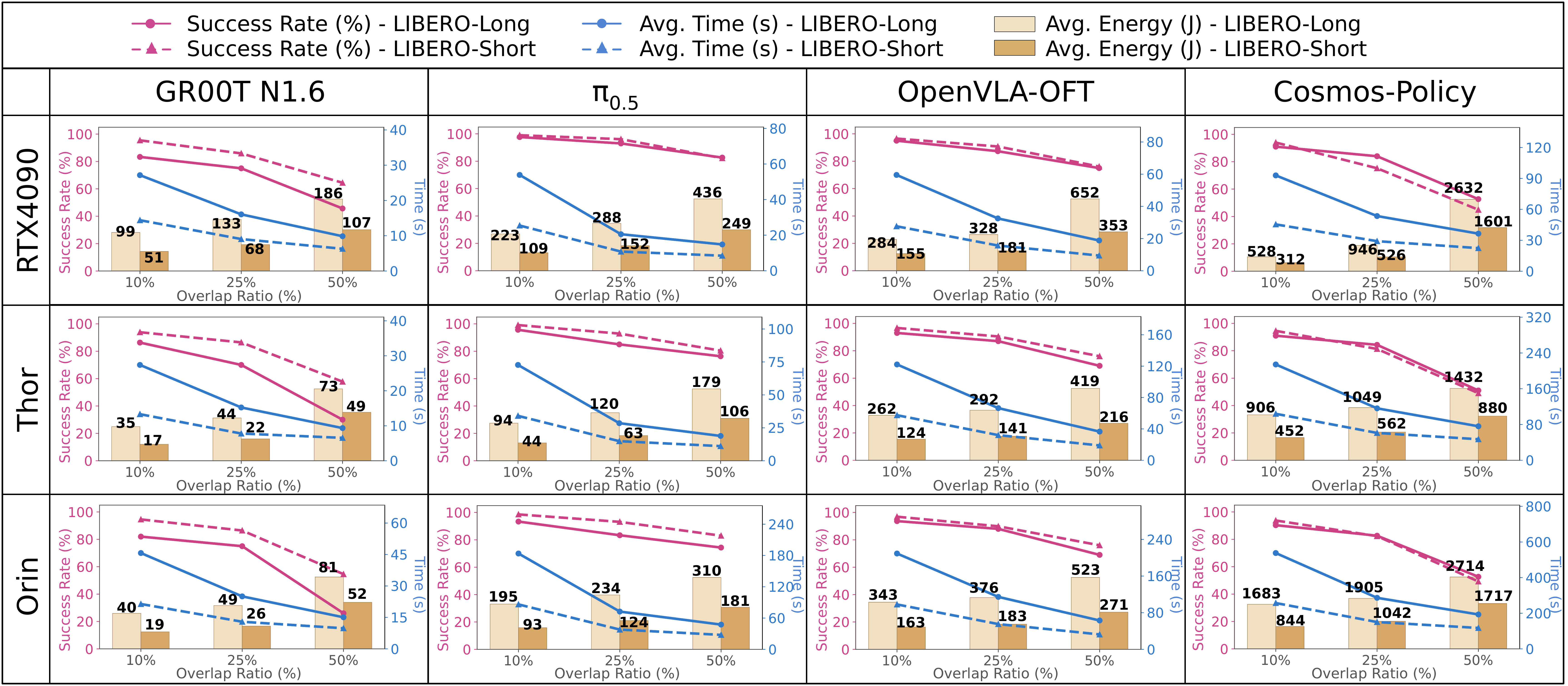}
\vspace{-0.25in}
\caption{Accuracy--speed--energy tradeoff across all model--platform combinations, split into
our two task groups: \emph{Long} (LIBERO-10) and \emph{Short} (LIBERO-Goal/Spatial/Object).
Each subplot shows success rate (magenta, left axis), average episode time (blue, right axis),
and average measured energy (tan bars, annotated in J) as a function of overlap percentage.}
\Description{A grid compares success rate, episode time, and measured energy across overlap levels for every model and platform, with separate curves for Long and Short task groups.}
\label{fig:tradeoff_grid}
\vspace{-0.05in}
\end{figure*}

\subsection{Experimental Setup}
\label{subsec:overlap_design}

\textbf{Overlap mechanism and control period.}
A chunk contains $K$ actions, and the controller executes one action per control period.
The key design parameter of asynchronous overlap is the \emph{overlap percentage}~$p$, which determines when the next inference begins~\cite{shukor2025smolvla, tang2025vlash, black2025real}.
The system triggers it when $\lceil K \times p \rceil$ steps remain in the current chunk, leaving that many control periods for the inference to complete before a stall occurs.
For $K{=}16$, $p{=}10\%$ provides two control periods for the next inference, whereas $p{=}50\%$ provides eight.
The per-step control period must therefore satisfy
\begin{equation}
\Delta t \;\geq\; \frac{T_{\text{infer}}}{\lceil K \times p \rceil},
\label{eq:dt_lower_bound}
\end{equation}
where $T_{\text{infer}}$ is inference latency.
Higher $p$ permits a shorter $\Delta t$ but increases observation staleness (Eq.~\eqref{eq:staleness}), exposing the accuracy--staleness tradeoff of Section~\ref{subsec:closed_loop}.

\textbf{Serving configuration.}
For closed-loop evaluation, the RTX\,4090 runs with default DVFS, while Jetson Thor
and Jetson AGX Orin run at maximum clocks. We use the RTT measured under these
settings to set the control period:
\begin{equation}
\Delta t = \frac{\text{RTT}_{\max} \times 1.5}{\lceil K \times p \rceil},
\label{eq:control_period}
\end{equation}
where $\text{RTT}_{\max}$ is each episode's maximum round-trip time for one inference
request, measured under synchronous operation and averaged across episodes, and the
1.5$\times$ factor adds headroom beyond the observed worst case.
Compilation follows Table~\ref{tab:compile_config}, per-configuration control periods are in Table~\ref{tab:control_period}, and the client--server system is shown in Figure~\ref{fig:system_architecture}.

\textbf{Evaluation scale.}
We sweep $p \in \{10\%, 25\%, 50\%\}$ and run 30 independent rollouts per task across four models, three platforms, and four LIBERO suites (10 tasks each), grouped by horizon into a \emph{Long} group (LIBERO-10) and a \emph{Short} group (LIBERO-Goal, LIBERO-Spatial, LIBERO-Object), totaling 43{,}200 closed-loop episodes over 40 tasks.

\subsection{The Accuracy--Speed--Energy Tradeoff}
\label{subsec:tradeoff}

Figure~\ref{fig:tradeoff_grid} presents the three-way tradeoff across all model--platform combinations for the Long and Short task groups.
Every configuration shows the same directional pattern as overlap grows from 10\% to 50\%: success rate decreases, time decreases, and energy increases. The rates, however, differ substantially.

\textbf{Accuracy degradation is architecture-dependent.}
The success-rate drop from 10\% to 50\% overlap maintains a consistent ordering across platforms \emph{and across both task groups}: $\pi_{0.5}$ degrades least, followed by OpenVLA-OFT, then GR00T N1.6 and Cosmos-Policy (on RTX\,4090 the two robust architectures lose $15$--$20$\,pp, versus $31$--$49$\,pp for GR00T N1.6 and Cosmos-Policy).

\textbf{Speed scaling is platform-invariant.}
For a fixed task group, the time reduction ratio from 10\% to 50\% is stable across platforms (e.g., GR00T N1.6 2.8--3.0$\times$; $\pi_{0.5}$ 3.6--3.9$\times$ on the Long group).
This follows from the stage-proportion stability observed in Section~\ref{subsec:soc_scaling}: because the backbone-to-action-head latency ratio is preserved across platforms and control periods are calibrated to the measured worst-case RTT, the relative speedup is platform-independent.

\textbf{Energy scaling is platform-dependent.}
Cosmos-Policy on RTX\,4090 exhibits the steepest energy growth (up to 5.1$\times$ in the Short group from 10\% to 50\%), while GR00T N1.6 on the same platform increases only 1.9--2.1$\times$. On edge SoCs, the range is 1.5--2.9$\times$ due to the lower power ceiling.

\begin{takeawaybox}[takeaway:async_tradeoff]
Asynchronous overlap creates a three-way tradeoff: accuracy degradation depends on
architecture, relative speedup is stable across platforms, and energy scaling depends on
platform.
\end{takeawaybox}

\subsection{Throughput--Energy Efficiency}
\label{subsec:throughput_energy}

\begin{figure}[t]
\centering
\includegraphics[width=\columnwidth]{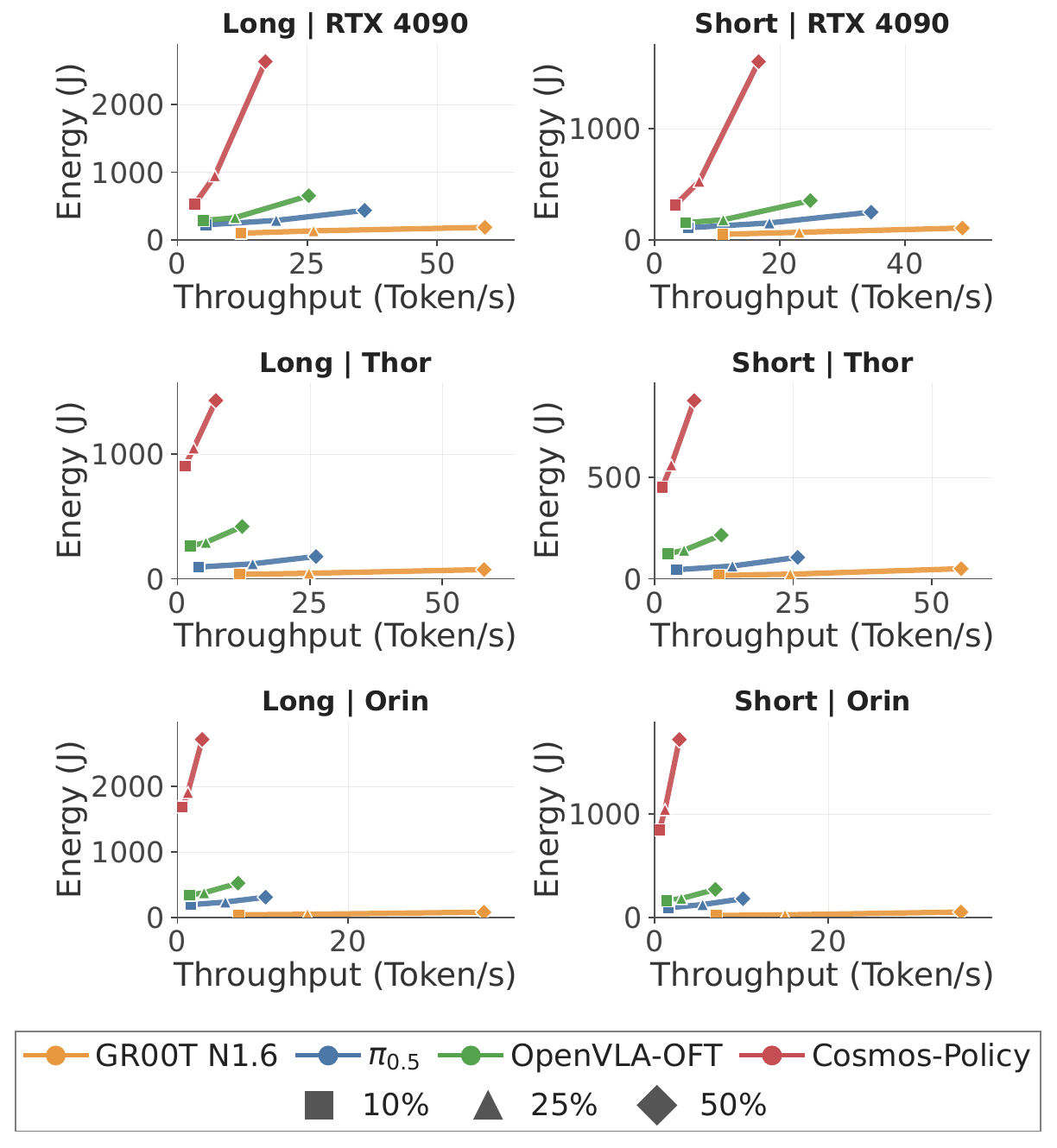}
\vspace{-0.2in}
\caption{Task-equal successful-episode energy versus action throughput, per task group
and platform, across overlap levels (10\%--50\%). Each trajectory connects one model's overlap
settings; lower right (higher throughput at lower energy) is better.}
\Description{Panels plot task-equal energy per successful episode against action throughput, with trajectories connecting each model's overlap settings on each platform and task group.}
\label{fig:throughput_energy}
\vspace{-0.1in}
\end{figure}

Figure~\ref{fig:throughput_energy} plots task-equal successful-episode energy against generated action throughput.
Each point represents one model--platform--overlap combination.
Each curve traces a model's trajectory as overlap increases from 10\% to 50\%, moving rightward (higher throughput) and upward (higher energy).

{\sloppy 
\textbf{Memory-bound stage dominance corresponds to higher throughput-per-joule.}
GR00T N1.6 and $\pi_{0.5}$ trace shallow curves across all platforms.
On RTX\,4090, GR00T N1.6 reaches ${\sim}$59\,tokens/s at 186\,J per successful episode; Cosmos-Policy reaches ${\sim}$17\,tokens/s at 2{,}632\,J.
On Thor, the same efficiency ordering holds: GR00T N1.6 achieves 58\,tokens/s at 73\,J, while Cosmos-Policy exceeds 1{,}000\,J despite far lower throughput.
\par}

{\sloppy
\textbf{Edge platforms show a larger efficiency gap.}
The energy gap between the most and least efficient models spans roughly one order of magnitude on RTX\,4090 and widens further on Thor.
GR00T N1.6 and $\pi_{0.5}$ keep their shallow energy curves as throughput increases, whereas Cosmos-Policy and OpenVLA-OFT retain steep slopes.
This ordering is consistent with the single-inference profiles in Section~\ref{subsec:roofline}, where memory-bound stages, concentrated in the action heads, account for most of GR00T N1.6's and $\pi_{0.5}$'s GPU time, while compute-bound stages dominate Cosmos-Policy and OpenVLA-OFT.
\par}

\begin{takeawaybox}[takeaway:throughput_per_joule]
The stage-level bottleneck mix identified by single-inference profiling
(Takeaway~\ref{takeaway:mem_compute_bound}) is reflected in deployment efficiency.
In the evaluated models, dominance by memory-bound action heads corresponds to
over an order of magnitude higher throughput-per-joule, and edge platforms
widen this gap.
\end{takeawaybox}

\subsection{Deployment Operating Points Under SLO Constraints}
\label{subsec:slo}

\begin{figure*}[t]
\centering
\includegraphics[width=\textwidth]{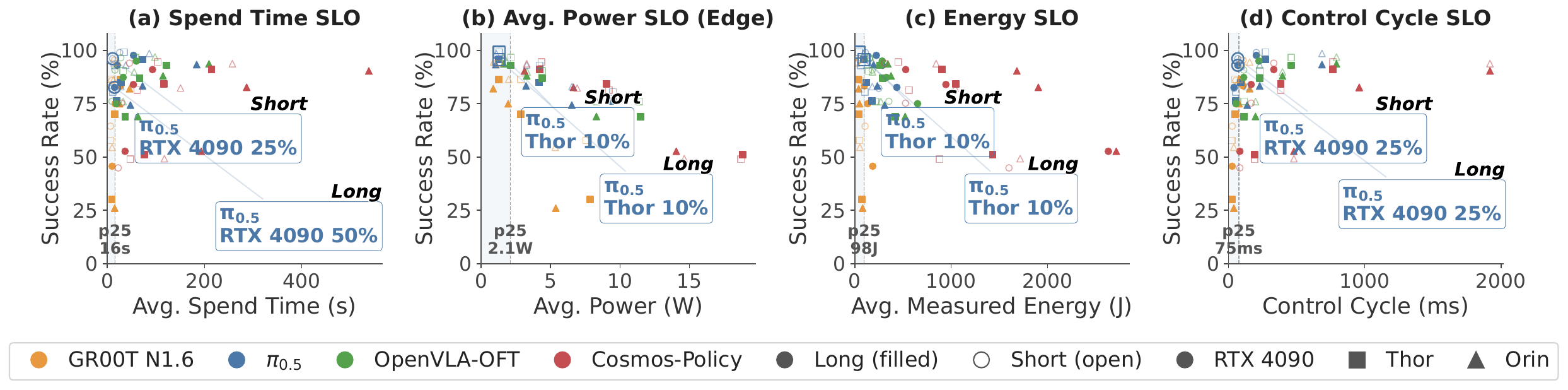}
\vspace{-0.3in}
\caption{Deployment operating points under SLO constraints.
Shading marks values below each metric's lower-quartile (p25) budget; callouts mark the highest-SR feasible Long and Short points.
Filled markers: Long (LIBERO-10); open markers: Short (LIBERO-Goal/Spatial/Object).}
\Description{Four scatter plots show success rate against episode time, edge power, measured energy, and control-cycle latency, with shaded feasible regions and selected Long and Short operating points.}
\label{fig:slo_scatter}
\vspace{-0.05in}
\end{figure*}

Figure~\ref{fig:slo_scatter} presents operating points under four resource
and timing constraints. We additionally evaluate an SR-first minimum-energy
objective. To apply a common selection rule, we set each budget to the lower
quartile (p25) of its measured configuration distribution and select the
highest-SR feasible point.

{\sloppy
$\pi_{0.5}$ is selected under all four p25 budgets, while the constraint
determines its platform and overlap. The 15.6\,s spend-time budget selects
RTX\,4090 at 50\% overlap on Long (82.7\% SR) and 25\% on Short (96.1\% SR).
The 75.2\,ms control-cycle budget also selects RTX\,4090, at 25\% overlap for
both groups (93.0\% and 96.1\% SR). The edge-only 2.1\,W power budget and the
98.0\,J energy budget select Thor at 10\% overlap for both groups (95.7\% and
99.1\% SR). Under the energy budget, the Long and Short selections consume
93.5\,J and 44.4\,J, respectively.
\par}

{\sloppy
The selected model changes with the objective: from highest SR within a
resource budget to minimum energy above an SR threshold. With an
85\% SR threshold, GR00T N1.6 on Thor at 10\% is selected for both
Long (86.3\%, 34.8\,J) and Short (93.9\%, 16.8\,J), although the Long result is
close to the threshold. OpenVLA-OFT and Cosmos-Policy are not selected under either
objective. These selections do not establish model-level dominance: the
budgets and the threshold are analysis choices, and different cut-offs could
select different configurations.
\par}

\begin{takeawaybox}[takeaway:no_pareto_dominant]
No single model--platform--overlap configuration is Pareto-dominant
across time, energy, power, and control-cycle constraints. The
optimal operating point shifts with the SLO type and task group, though
$\pi_{0.5}$ is selected under most SLOs.
\end{takeawaybox}

\section{Discussion}
\label{sec:discussion}

\textbf{Staleness-aware VLA serving abstraction.}
Our measurements motivate a runtime interface that exposes four kinds of state: per-stage
cost, action-chunk progress, observation freshness, and deployment SLOs. Concretely, the
interface reports each stage's latency and compute/bandwidth demand, the number of
unexecuted actions in the current chunk and the time at which they run out, the timestamp
and staleness budget of the latest observation, and the target SLOs for latency, energy, power, and task success. A
scheduler could use this state to select execution timing, overlap ratio, resource settings,
and where each stage executes. LLM prefill--decode disaggregation schedules worker pools
across concurrent requests~\cite{zhong2024distserve}. A batch-1 VLA loop instead schedules
stages within one robot's inference, where the execution location of a stage changes
observation staleness and task success in addition to latency
(Section~\ref{sec:closed_loop_inference}).

\textbf{Architecture implications.}
The measured bottlenecks suggest several hardware features that could improve VLA
inference, although we do not propose a specific accelerator design. Large GEMMs in
backbones and video-generation DiTs benefit from high tensor-compute throughput. Denoising
loops instead issue many small, fixed-shape GEMMs, so they benefit from hardware that
sustains high utilization on small matrix operations. Memory-bound action heads benefit from
higher memory bandwidth, and iteration-heavy action heads benefit from low kernel-launch and
control overhead. A denoising loop also reloads the same weights at every step, so cache or
local memory that keeps them on chip could reduce this repeated traffic. Finally, faster
frequency transitions would reduce the measured overhead that limits our phase-aware DVFS
policy.

{\sloppy
\textbf{Selective applicability of existing optimizations.}
The control-period deadline favors on-board or nearby edge execution because remote
execution adds network latency and jitter to the observation-staleness budget
(Section~\ref{sec:closed_loop_inference}). In the single-robot setting studied here,
requests arrive one at a time, so inference runs at batch size one. We assess the
optimizations discussed below using three workload properties: request concurrency,
autoregressive generation, and the dominant execution stage.
\par
Techniques that optimize computation within a single inference (kernel fusion and graph
capture, step distillation, feature caching, weight-only quantization) remain applicable at
batch size one, although their benefit depends on the dominant stage. Among these, we
directly evaluate compilation, which delivers 4.4--7.4$\times$ speedups on iteration-heavy
action heads (Section~\ref{subsec:stage_compilation}).
\par
Request-level and KV-cache optimizations are more constrained in this setting. Because a
single robot issues only one request at a time, continuous batching offers little benefit
unless a server handles requests from multiple robots. The autoregressive baseline still
uses a KV cache, but its decode length is short and fixed by the action format, so it is
known before decoding starts. This leaves little opportunity for paged KV-cache management
to reduce fragmentation or improve memory utilization. The cache is also discarded after
each action is generated, so the profiled implementation retains no prefix to reuse across
inferences (Section~\ref{subsec:embodied_models}).
\par
Techniques specific to autoregressive decoding, such as speculative decoding, have a still
narrower scope. Among the models we study, they apply only to the autoregressive baseline.
Even there, the decode is so short that the overhead of running a draft model may outweigh
the benefit. Except for compilation, these conclusions are based on workload structure
rather than measured speedups.
\par}

\textbf{Scope and limitations.}
\emph{System-design scope.} We characterize the constraints a VLA runtime must handle,
but we do not implement a scheduler. A deployable phase-aware policy must account for transition
overhead and phase duration. Stage co-execution must account for model dependencies and
shared-memory contention.

\emph{Simulation scope.} Across 40 LIBERO tasks, we compare the models under identical MuJoCo
dynamics in 43{,}200 episodes spanning model, platform, and overlap configurations.
MuJoCo provides controller-mediated actuation dynamics, inference timing comes from the target
hardware, and each control period includes a conservative 1.5$\times$ margin over the measured worst-case RTT. This setup
relates control frequency and observation staleness to task success under controlled conditions.
It does not replace real-robot or hardware-in-the-loop validation: sensor pipelines, network
jitter, multi-camera synchronization, unmodeled contact dynamics, and physical disturbances are
outside our scope.

\emph{Model scope.}
We profile autoregressive OpenVLA as the discrete action-token baseline. The closed-loop
evaluation covers the four action-chunk models used for continuous high-rate control. The
OpenVLA profile retains the known memory-bound LLM decode behavior. Alternative
autoregressive control interfaces are outside our scope.

Takeaway~\ref{takeaway:mem_compute_bound} may extend to other models in the same paradigm,
although exact stage proportions depend on model size and chunk length. Our model set excludes
1-step distillation variants, which can remove or reduce the memory-bound action-head stage.
Whether the accuracy--staleness ordering (Takeaway~\ref{takeaway:async_tradeoff}) holds for such
architectures remains open.

\emph{Hardware scope.} We evaluate one discrete edge server and two unified-memory onboard
SoCs, all from NVIDIA. Sensitivity varies across them with compute-to-bandwidth balance.
Non-NVIDIA accelerators are not evaluated.

\emph{Energy-measurement scope.} Cross-platform energy comparisons reflect relative trends
rather than absolute ratios due to differing measurement boundaries
(Section~\ref{sec:methodology}), while within-platform comparisons are directly comparable.

\section{Related Work}
\label{sec:related_work}
\textbf{Algorithmic perspectives on VLA efficiency.}
Recent surveys organize the VLA landscape along
algorithmic axes: Kawaharazuka~et~al.~\cite{kawaharazuka2025vla} 
provide a full-stack review spanning architectures and robot platforms, 
Yu~et~al.~\cite{yu2026survey} propose a taxonomy of efficient VLAs 
across model design, training, and data collection, and 
Guan~et~al.~\cite{guan2025efficient} categorize efficiency techniques 
by architecture, perception, action generation, and training/inference 
strategy. These works describe proposed methods but
provide limited empirical characterization of how deployed systems
behave on specific hardware. Deployment-level bottleneck attribution,
DVFS-sensitive energy behavior, and closed-loop operating-point
analysis receive less empirical attention.

\textbf{LLM serving systems.}
LLM servers improve throughput and goodput for concurrent autoregressive
requests through scheduling and
batching~\cite{yu2022orca, agrawal2024taming}, paged KV-cache
management~\cite{kwon2023efficient}, and prefill--decode
disaggregation~\cite{zhong2024distserve, patel2024splitwise}.
VLA serving changes both the scheduling unit and the objective. A single-robot
batch-1 loop schedules backbone and action-head stages instead of prefill and
decode phases across independent requests. Placement and queuing delay also
increase observation staleness, coupling task success to control-period,
accuracy, and energy constraints.

\textbf{ML inference characterization.}
Roofline models explain memory-bound LLM decoding and guide serving
optimization~\cite{yuan2024llm}. RooflineBench~\cite{bi2026rooflinebench}
extends operational-intensity comparison to on-device LLMs. Roofline analysis
also explains the memory-bound autoregressive OpenVLA decode loop in
Section~\ref{subsec:embodied_models}. We apply it to heterogeneous action-chunk
computation across iterative denoising, parallel decoding, and video
generation, and separately measure closed-loop staleness in batch-1 serving.

\section{Conclusion}
This work presents a system-level workload characterization of embodied action generation model serving across four models, three platforms, and 40 LIBERO tasks. At the single-inference level, we show that the compute- versus memory-bound behavior is largely governed by action tensor dimensionality. Under closed-loop overlap, accuracy degradation depends on model architecture, while energy scaling depends on the hardware platform, with no configuration remaining Pareto-dominant across SLOs. These results highlight the need for architecture-aware resource allocation and staleness-aware operating-point selection. Together, our findings call for VLA systems that jointly consider model architecture, hardware resource balance, and runtime policies, providing a foundation for future accelerators and serving systems tailored to embodied AI.

\begin{acks}
This work was supported by the Institute of Information \& Communications Technology Planning \& Evaluation (IITP) funded by the Ministry of Science and ICT, Korea (RS-2024-00396013 and RS-2024-00402898), and the National Research Foundation of Korea (NRF; RS-2024-00347114). 
\end{acks}

\bibliographystyle{ACM-Reference-Format}
\balance
\bibliography{refs}

\end{document}